\documentclass[lettersize,journal]{IEEEtran}
\usepackage{amsmath,amsfonts}
\usepackage{algorithmic}
\usepackage{algorithm}
\usepackage{array}
\usepackage[caption=false,font=normalsize,labelfont=sf,textfont=sf]{subfig}
\usepackage{textcomp}
\usepackage{stfloats}
\usepackage{url}
\usepackage{verbatim}
\usepackage{graphicx}
\usepackage{cite}
\usepackage{hyperref}

\begin{document}

\title{Evaluation of Spherical Wavelet Framework \\
in Comparsion with Ambisonics}

\author{Şafak Ekmen, Hyunkook Lee
\thanks{Ş. Ekmen and H. Lee are with Applied Psychoacoustics Laboratory (APL), University of Huddersfield, Huddersfield, UK}
}

\markboth{Preprint}
{Ekmen, Lee, \MakeLowercase{\textit{et al.}}: Evaluation of Spherical Wavelet Framework
in Comparsion with Ambisonics}


\maketitle

\begin{abstract}
Recently, the Spherical Wavelet Framework (SWF) was proposed to combine the benefits of Ambisonics and Object-Based Audio (OBA) by utilising highly localised basis functions. SWF can enhance the sweet-spot area and reduce localisation blur while still enabling a sparse representation of the complete sound field, making storage and transmission more efficient. Initial vector analysis and listening test of SWF have shown promising results; however, these findings are limited to very specific conditions and do not include perceptual metrics. The present study investigates SWF in greater detail, comparing it with Ambisonics. The comparison was carried out using IACC, ITD, and ILD estimations, as well as listening tests with ecologically valid sound sources. Various reproduction layouts: regular polyhedron, t-design, and Lebedev grid with their corresponding Ambisonics orders and channel counts were evaluated. Results indicate that SWF is rated significantly more similar to the reference than Ambisonics is, in terms of overall spatial and timbral fidelity; however, it is considerably dependent on the subdivison of the sphere. Moreover, it cannot natively represent a wave arriving at a continuous direction. Possible solutions are proposed.
\end{abstract}


\section{Introduction}
\IEEEPARstart{A}{mbisonics} is the most practical technique for capturing, transmitting, storing and reproducing sound-fields. It is essentially the Fourier Transform on a sphere, referred to as the Spherical Harmonic Transform (SHT). Using SHT, Ambisonics can represent an acoustic field by a linear combination of basis functions that are called Spherical Harmonics (SH). This representation encodes a sound-field into the Spherical Harmonic Domain (SHD) either through microphone capsules placed at points that sample the sphere or through signal processing techniques. This representation, in turn, enables efficient storage, transmission and rendering of the sound-field and, it is theoretically layout independent. 

However, Ambisonics requires large number of SHs and therefore channel counts to provide a high-quality reproduction. Moreover, mainly because SHs are continuous at everywhere on the sphere, Ambisonics introduces inter-channel crosstalk in reproduction. In return, this inter-channel crosstalk may distort perceptual cues and does not allow for a large sweet-spot \cite{leeSpatialTimbralFidelities2019a}.  This accompanied by practical reasons such as ability to choose different microphone capsules, and the complexity of the processing with Higher Order Ambisonics, it did not gain much traction, especially in commercial music and film applications. That is being said, its efficiency, particularly for transmission and head-tracking makes it the preferred technique for applications such as streaming and VR \cite{leeSpatialTimbralFidelities2019a}.

In commercial music and film applications, Object-Based Audio (OBA) frameworks are dominating. OBA represents each or group of sound sources with metadata as objects. This metadata contains information about the sound source(s) it is attached to, such as panning and gain information. Usually, these objects are rendered at the reproduction end based on the metadata. OBA formats, arguably, are easier to implement into the traditional production workflows. However, they are layout dependent, have proprietary variations and computationally expensive. Also, OBA formats do not provide a complete representation of a sound-field.

The above shortcomings of SHT are also encountered when the Fourier Transform is applied to the line (time and frequency domain). A common solution is to use Wavelet Transform that has compactly supported basis functions referred as Wavelets and by using these functions, Wavelet Transform provides a tuneable and variable-resolution representation that is better localized either in frequency or in time. By using localised basis functions, SWT allows for sparse approximation of a function on sphere. In audio applications this may provide advantages such as lower channel counts, decreased localisation blur, and increased sweet spot along with more flexible frequency dependent processing compared to Ambisonics. 

Spherical Wavelet Transform (SWT) has been in use in geodesy \cite{chevrotOptimizedDiscreteWavelet2012}, astronomy \cite{barreiroTestingGaussianityCOBE2000} and computer graphics fields for some time now. However, to the best of the authors' knowledge, its applications in acoustics are rare. Kensuke \cite{kensukeOptimalHaarWavelet2018}, applied SWT for decomposition of acoustic fields. Recently, Scaini \cite{scainiWaveletbasedSpatialAudio2019} applied SWT to spatial audio using a particular transform algorithm and wavelet families based on \cite{schroderSphericalWaveletsEfficiently1995} and named it as Spherical Wavelet Framework (SWF). Later, studies in \cite{eguinoaSubjectiveEvaluationLocalization2021} and \cite{narvaezAdvancingWaveletBasedSpatial2022} further investigated and developed onto the SWF, respectively. Evaluation of these studies shows promising results in favour of SWF. However, they are conducted for  limited conditions such as considering only one type of loudspeaker array or only conducting vectorial analysis while not considering perceptually oriented metrics such as Inter-aural Cross Correlation (IACC), Inter-aural Time Difference (ITD), Inter-aural Level Difference (ILD) or spectrum fidelity.

This paper investigates spherical wavelets for spatial audio and specifically SWF in greater detail, considering its usability for  encoding, processing, and reproduction  of spatial audio. Both objective analysis and subjective listening test were performed. As objective analysis, first, auditory modelling is employed in which IACC, ITD, ILD were calculated using related auditory models. Their errors with respect to a reference were computed. In addition to these, to assess the spectral fidelity, Perceptual Spectral Difference (PSD) as explained in \cite{armstrongPerceptualSpectralDifference2018, mckenziePredictingColourationBinaural2022} is calculated. This followed by vectorial analysis. In vectorial analysis, energy (rE) and (rV) velocity vectors were calculated along with different spread metrics that are previously proposed in the literature \cite{danielRepresentationChampsAcoustiques2000, epainAmbisonicDecodingConstant2014, frankSourceWidthFrontal2013}. After objective analysis, a subjective listening test based on the Multiple Stimuli Hidden Reference and Anchor (MUSHRA) procedure was conducted both on spatial and tonal fidelity.

The rest of the paper is structured as follows. In Theoretical Background section, a brief overview of the theory of Spherical Wavelet Transform is presented by building onto Spherical Harmonic Transform and summarizing its historical development. At the last part of the section, known applications of SWT to acoustic fields are reviewed. Conducted analyses and tests are presented separately in their own sections. Each of these sections are devided into Methodology and Results subsections. Detailed explanations of used metrics, algorithms and procedures are placed in the Methodology subsections. Results and their analysis are presented in the Results subsections. It is followed by Discussion and Further Study section and the paper is aptly finished with the  conclusion.

\section{Theoretical Background}
Sound pressure $p$ from a far-field source on a point $r$ on an acoustically transparent sphere can be expressed by the homogenous, lossless wave equation \cite{politisMicrophoneArrayProcessing2016, rafaelyAcousticalBackground2019}:
\begin{equation*}
(\nabla^2 - \frac{1}{c^2}\frac{\partial^2}{\partial^2t})p(r, t) = 0
\end{equation*}
where $\nabla^2$ is the Laplacian and $c$ is the speed of sound. This transparent sphere is mathematically referred as 2-sphere which is a manifold in $\mathbb{S}^2$. Conceptually a 2-sphere can be imagined as a sphere that is obtained by folding a massless sheet in space that does not show any resistance to any type of energy. In the frequency domain, wave equation becomes the Helmholtz equation \cite{politisMicrophoneArrayProcessing2016}:
\begin{equation*}
(\nabla^2 + k^2)p(r, \omega) = 0
\end{equation*}
where $k = \omega/c$ is the wave number and $\omega$ is the frequency. Using Helmholtz equation, sound-fields can be decomposed to continuous number of plane-waves each with amplitude $a(\gamma)$ where $\gamma$ is the direction of the plane-wave arriving at the point r \cite{politisMicrophoneArrayProcessing2016}:
\begin{equation}
\int_{\mathbb{S}^2}a(\gamma, \omega)\psi_{k,\gamma}d\gamma
\label{eq3}
\end{equation}
Here, $\psi_{k,\gamma}$ is the basis function that with different $k$ and $\gamma$ values, constitutes the components of the decomposed sound-field. If the basis functions are Legendre functions forming an orthonormal basis, then this is called Spherical Harmonic Transform \cite{jarettSphericalArrayProcessing2014} on 2-sphere. Continuous SHT can be thought as the analytic solution to Helmholtz equation. Ideally, to get the exact representation of the sound pressure on the surface of the sphere one must use infinite number of spherical harmonics. However, this is not possible in real-world since it would require infinite number of sensors as well as being not possible to compute. Therefore, a discretization of the sphere is necessary \cite{jarettSphericalArrayProcessing2014}. This is referred as Discrete Spherical Harmonic Transform, and it is the numerical solution. As an example, in initial iteration of Ambisonics the tetrahedral microphone uses a tetrahedron to sample the sphere \cite{gerzonPeriphonyWithHeightSound1973, gerzonDesignPreciselyCoincident1975, cravenCoincidentMicrophoneSimulation1977a}. Using platonic solids for sampling is a known approach \cite{corneliusAnalysisSamplingGrids2015, elkoPolyhedralAudioSystem2015}. There are other sampling approaches which are categorized by the distance between the sampling points being equal which is referred as uniform sampling \cite{driscollComputingFourierTransforms1994} or being near-equal (near-uniform) \cite{corneliusAnalysisSamplingGrids2015, meyerSphericalMicrophoneArrays2004}. To preserve the orthonormality of the Spherical Harmonics, the sampling scheme must be uniform or at least near-uniform. In the near-uniform case, even an infinite number of samples are used, reconstruction of the sound-field will not be a perfect one but an approximation.

Basis functions $\psi_{k,\gamma}$ used for the SHT in Eq. \ref{eq3} are not compactly supported. This feature limits the ability to represent a sound-field that is localized (directional). It has been found that by using basis functions that are similar to the investigated phenomenon, as long as the basis functions forms an orthogonal space, a more sparse representation can be achieved \cite{antoineWaveletTransformManifolds2010b}. This idea is the foundation of the wavelet transforms.
Early work on the Spherical Wavelet Transform (SWT) focused on defining continuous transforms by adapting the dilation operator from the plane to the sphere, often using projection-based methods. Among these, group-theoretic approaches resemble Spherical Harmonics, as they define operations and basis functions directly on the sphere. This makes them naturally suitable for spatial audio applications, due to their built-in rotation groups that simplify sound field manipulation.

Torresani \cite{torresaniPositionfrequencyAnalyisSignals1995b} proposed a Continuous SWT (CSWT) on a homogeneous space-frequency domain, where spherical rotations act as translations and modulations serve as frequency band. Using a tangent phasor to define the frequency scales, the transform can be discretized.

Antoine and Vandergheynst \cite{antoineWaveletsNsphereRelated1998, antoineWavelets2SphereGroupTheoretical1999} used a group-theoretic approach combined with stereographic projection to define translation and dilation operators, allowing for discretization and a rotatable wavelet basis. This remains a key example of projection-based SWT design.

In contrast, Freeden and Windheuser \cite{freedenSphericalWaveletTransform1996} offer a non-projection-based approach, developing scale-discretized wavelets on the harmonic space using singular integral theory. Their wavelets are defined in harmonic space and support multiresolution analysis (MRA) with square-integrable basis functions.

Schröder and Sweldens \cite{schroderSphericalWaveletsEfficiently1995} also avoid defining dilation by using lifting scheme on the subdivisions of Platonic solids to approximate the sphere. Their lifting scheme allows a computationally efficient discrete transform that can work with arbitrary meshes. However, it lacks an explicit rotation group and requires interpolation for arbitrary directions.

Building on Antoine and Vandergheynst’s work, Bogdanova et al. \cite{bogdanovaStereographicWaveletFrames2005} introduced half-continuous and discrete wavelet frames, combining basis functions with discrete sampling intervals. This enables discrete transforms, rotation groups, and MRA, although interpolation between levels may introduce artifacts depending on the method.

Wiaux et al. \cite{wiauxExactReconstructionDirectional2008}, inspired by both Antoine and Vandergheynst, and Holschneider \cite{holschneiderContinuousWaveletTransforms1996a}, proposed a Discrete Spherical Wavelet Transform (DSWT) that is natively defined on the sphere without projection. Their approach includes a dilation parameter and supports flexible MRA. They introduced a “harmonic dilation” operator defined in the harmonic domain, and their framework supports various wavelet families entirely within the spherical space.

Applications of SWT in audio are relatively rare. Kensuke \cite{kensukeOptimalHaarWavelet2018} applied SWT with triangular meshes and Haar wavelets for sparse acoustic field representation. Simulations showed smoother wavelets may perform better, especially at high frequencies, but the study lacked psychoacoustic metrics and used many more coefficients than typical audio applications.

Scaini \cite{scainiWaveletbasedSpatialAudio2019} and Scaini \& Arteaga \cite{scainiWaveletBasedSpatialAudio2020} proposed the Spherical Wavelet Framework (SWF) based on the Schröder and Sweldens approach. Using Loop Subdivision of an octahedron, they tested 6, 18, and 66-channel configurations. Continuous source positions were interpolated using tri-linear interpolation, then wavelet coefficients were mapped to a coarser resolution and decoded using a custom method ("IDHOA"). Evaluation using acoustic intensity and velocity vectors showed that even the 6-channel SWF performed comparably to 3rd-order Ambisonics (16 channels), though the latter produced more coherent sound fields. The study noted that the lifting scheme may not yield wavelets smooth enough for audio, leading to "jumpiness" in panning, and lacked psychoacoustic or formal listening tests.

Eguinoa et al. \cite{eguinoaSubjectiveEvaluationLocalization2021} conducted MUSHRA tests comparing SWF and Ambisonics in terms of localization and source width. Using pink noise and a 24-loudspeaker t-design setup, SWF level-1 was rated the highest, with level-0 matching 3rd-order Ambisonics. However, the study used synthetic stimuli and a speaker layout not derived from SWF’s geometry, so results depended heavily on decoder performance.

Narvaez \cite{narvaezAdvancingWaveletBasedSpatial2022} extended SWF to irregular loudspeaker arrays, shifting optimization from a dedicated decoder to the wavelet decomposition itself. Evaluation with acoustic estimators confirmed similar performance to previous SWF studies, but adapted to irregular setups.

\section{Objective Evaluation}
Objective evaluation of Spherical Wavelet Framework is performed in comparison to Ambisonics by auditory parameter analysis and vector analysis. In auditory parameter analysis, IACC, ITD, ILD and PSD values were calculated for different SWF and Ambisonics reproductions and compared to a direct-HRTF rendered reference. For comparison, errors were calculated as Mean Signed Differences (MSD) between the rendering technique and the reference for IACC, ITD and ILD metrics. Perceptual Spectral Difference is the difference with respect to a reference and therefore computed as such. 

For vectorial analysis, magnitudes of rE and rV vectors resulting from the simulated loudspeaker reproduction were calculated. This is accompanied by a spread metric that is based on the rE vector. Simulated values from SWF and Ambisonics were compared to each other.
\subsection{Auditory Modelling}
\subsubsection{Methodology}
Auditory parameters used in the objective evaluation are explained as follows. IACC metric measures the cross-correlation between the ear signals. It is commonly accepted as the indication of perceived source width \cite{lee3DMicrophoneArray2021}. Additionally the IACC error between the reference and a tested system may indicate distortion in frequency spectrum since the IACC of the reference for a particular position shows the ideal phase relationship between the ear signals. ITD and ILD metrics are most useful for localisation information on the horizontal plane \cite{macphersonListenerWeightingCues2002}. However, they contain insufficient information in terms of the perceived spectrum and of the localisation on the median plane which is heavily affected by the spectrum \cite{leeInvestigationPhantomImage2015}. 

To investigate the spectral fidelity and reproduction accuracy on the median plane, Perceptual Spectral Difference (PSD) metric was employed. This metric is based on the study in \cite{armstrongPerceptualSpectralDifference2018} and proposed by \cite{mckenziePredictingColourationBinaural2022}. It essentially is the magnitude difference between the spectra of the corresponding ear signals however, ear signal spectra are weighted according to ISO 226 equal loudness contours and to the ERB scale. Magnitude of each frequency sample are converted from sones to the Phons scale. These modifications are reported to better model the reaction of human auditory system to loudness and frequency \cite{armstrongPerceptualSpectralDifference2018, mckenziePredictingColourationBinaural2022}. 

IACC and ITD values of the SWF and Ambisonics renderings were calculated using the MaxIACCe-LP method which is reported to provide reliable results \cite{katzComparativeStudyInteraural2014a}. In this method, first, left and right ear signals had been low-passed at 3 kHz using a 5th-order Butterworth filter. Then, cross-correlation of the Hilbert envelops of the ear signals were calculated for the time-lags between – 1ms to 1ms. The time-point of the maximum cross-correlation coefficient was taken as the ITD value as in Eq. \ref{eq5} while the maximum cross-correlation coefficient itself was taken as the IACC. Implementation of this process is formulated in Eq. \ref{eq4} - \ref{eq6} where, $\tau$ is the time delay between the signals, $IACF(\tau)$ is the normalized interaural cross-correlation at the time delay $\tau$, $s_L$ and $s_R$ are the envelopes of the left and right ear signals respectively. 
\begin{multline}
IACF(\tau) =
\frac{\int_{0}^{t_{end}} s_L(t) s_R(t+\tau)\, dt}
     {\sqrt{\int_{0}^{t_{end}} s_L^2(t)\, dt
            \int_{0}^{t_{end}} s_R^2(t)\, dt}}, \\
-1\,\text{ms} < \tau < 1\,\text{ms}
\label{eq4}
\end{multline}
\begin{equation}
IACC = \max \left| IACF(\tau) \right|
\label{eq5}
\end{equation}
\begin{equation}
ITD=argmax \left| IACF(\tau) \right|
\label{eq6}
\end{equation}
ILD values were calculated as follows: left and right ear signals each had been filtered through 42 bands Gammatone Filter-Bank (GTFB). When the audible frequency range of 20 Hz to 20 kHz, is partitioned to 42 bands, each band corresponds to 1 Equivalent Rectangular Bandwidth (ERB). Then, rms of the each signal from each of the 42-channel output of the GTFB was calculated. ILD values for each ERB were calculated as the logarithmic ratio of the rms of the left and right ear signals as in Eq. \ref{eq7} where $s_L$ and $s_R$ are the left and right ear signals and $rms(.)$ is the root-mean-square. To obtain a single value, ILD values of the bands above 1.5 kHz were averaged. The 1.5 kHz point was decided based on the Duplex Theory \cite{macphersonListenerWeightingCues2002}, which states that, for human auditory localisation, ITD is most effective under 1.5 kHz while ILD is most effective above that frequency.
\begin{equation}
ILD=20\log_{10}(\frac{rms(s_L)}{rms(s_R)})
\label{eq7}
\end{equation}
PSD values are calculated using the “mckenzie2022” model from AMT \cite{majdakAMT1xToolbox2022}. Loudness normalisation and solid angle weighting was not applied since the samples were already amplitude normalised while they were compared for each position separately (e.g. SWF at (0, 0) against Ambisonics at (0, 0)). 

Computer generated, 100ms long pink noise was used as the programme material for the simulations. For the reference, input signals were convolved with the HRTFs belonging to the intended source position. Ambisonics samples were rendered in MATLAB environment using Politis’s HOA Library \cite{politisPolarchHigherOrderAmbisonics2025}. Input signal was encoded with the most suitable Ambisonics order for the target layout. Encoded signals are then decoded with the same order, using dual-decoder approach with the cut-off frequency of 800 Hz. Pseudo-inverse decoder was used for below the cut-off and max-rE decoder was used above \cite{hellerDESIGNIMPLEMENTATIONFILTERS2018, bertetInvestigationLocalisationAccuracy2013a}. First-order, third-order, fifth-order were used for Octahedron, 24-point t-design and 50-point Lebedev layouts respectively, For Ambisonics-MagLS rendering, mono input signal were encoded to Fifth-Order Ambisonics (5OA) using HOA Library, encoded signals are then decoded to headphones using SPARTA-AmbiBIN plug-in \cite{SPARTA2021} in Reaper DAW.  SWF samples were rendered in Python using Narvaez’s SWF package \cite{narvaezAdvancingWaveletBasedSpatial2022, narvaezSphericalWaveletFormat2022}. Simulated loudspeaker layouts are presented in Table \ref{table1}. Simulated panning positions are presented in Table \ref{table2}. Direct HRTF rendering for the reference was performed in Python using Sofar \cite{SofarSofar} and Pyfar \cite{PyfarPyfar} packages. 

Binaural renderings of the loudspeaker signals of SWF and Ambisonics were performed using so-called virtual loudspeaker approach. In this approach, loudspeaker signals that had been obtained from the rendering technique, were convolved with the HRTFs of the loudspeaker positions. Outputs of the convolution for each loudspeaker were then summed together for each ear. KU100 HRTFs from SADIE II Database \cite{armstrongPerceptualEvaluationIndividual2018a} were used. This database includes HRTF measurements for 8802 points which also contains the positions of loudspeakers in the layouts used in this study. In addition to these HRTF measurements, database also provides pseudo-inverse and max-rE decoding matrices for decoding Ambisonics to the used layouts. Based on these features, SADIE II Database is deemed to be suitable for the needs of this study. All the binaural renders were amplitude normalized to the 1 dBFS.
\begin{table}[!ht]
\caption{Tested Techniques and Layouts}\label{table1}
\centering
\begin{tabular}{|c||c|}
  \hline
  Technique & Reproduction Layout \\ 
  \hline
  Ambisonics & Octahedron \\
  \hline
  Ambisonics & 24-point t-design \\
  \hline
  Ambisonics & 50-point Lebedev \\
  \hline
  Ambisonics & MagLS \\
  \hline
  SWF & Octahedron \\
  \hline
  SWF & 24-point t-design \\
  \hline
  SWF & 50-point Lebedev \\
  \hline
\end{tabular}
\end{table}
\begin{table}[!ht]
\caption{Tested Panning Positions}\label{table2}
\centering
\begin{tabular}{|c||c|}
  \hline
  Azimuth $[^{\circ}]$ & Elevation $[^{\circ}]$ \\ 
  \hline
  0 & 0 \\
  \hline
  30 & 0 \\
  \hline
  90 & 0 \\
  \hline
  135 & 0 \\
  \hline
  180 & 0 \\
  \hline
  0 & 45 \\
  \hline
  0 & 90 \\
  \hline
  0 & 135 \\
  \hline
  45 & 45 \\
  \hline
  135 & 45 \\
  \hline
\end{tabular}
\end{table}
\subsubsection{Results}
\begin{figure*}[!ht]
    \centering
    \subfloat[]{\includegraphics[width=2in]{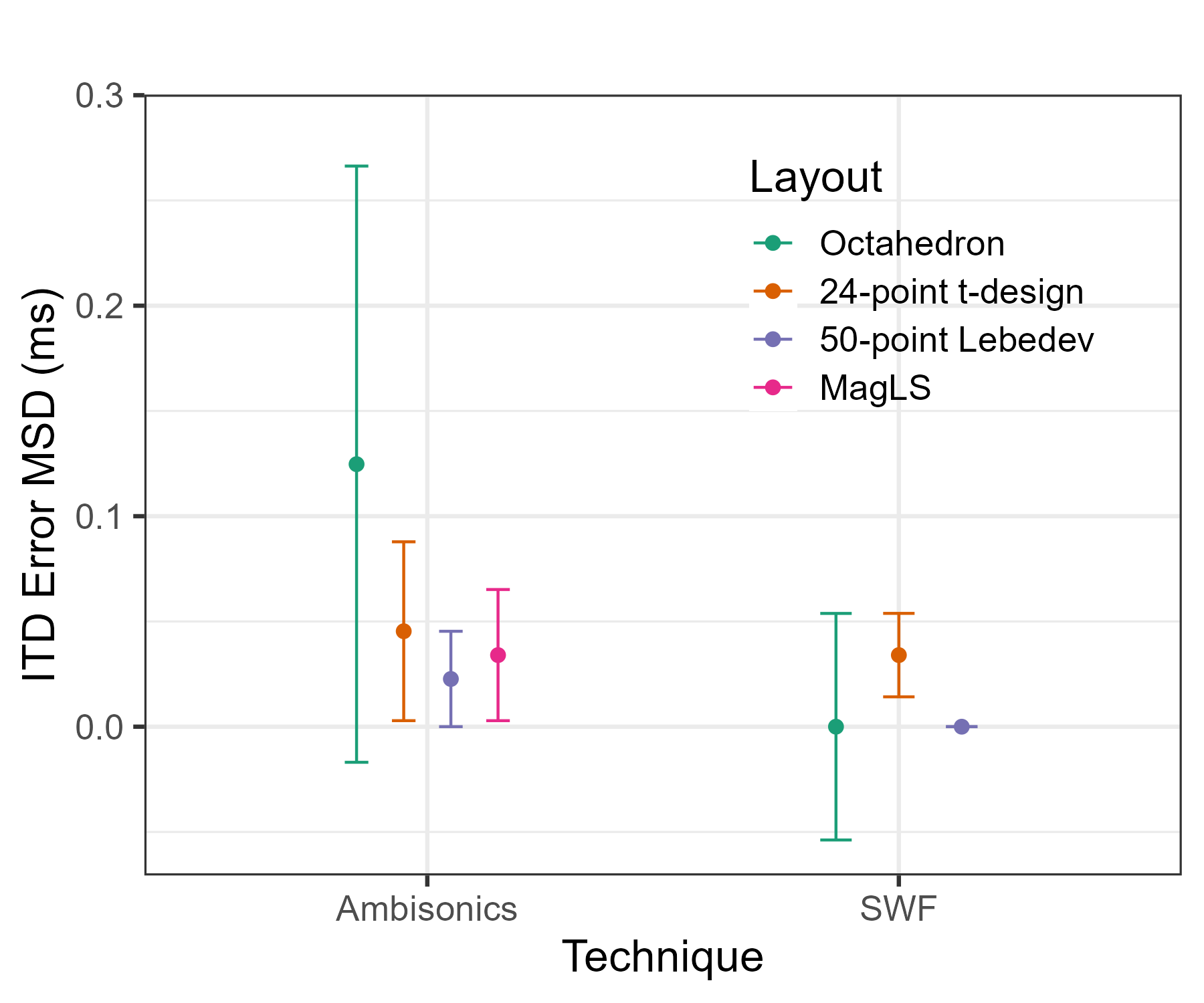}
    \label{fig1a}}
    \subfloat[]{\includegraphics[width=2in]{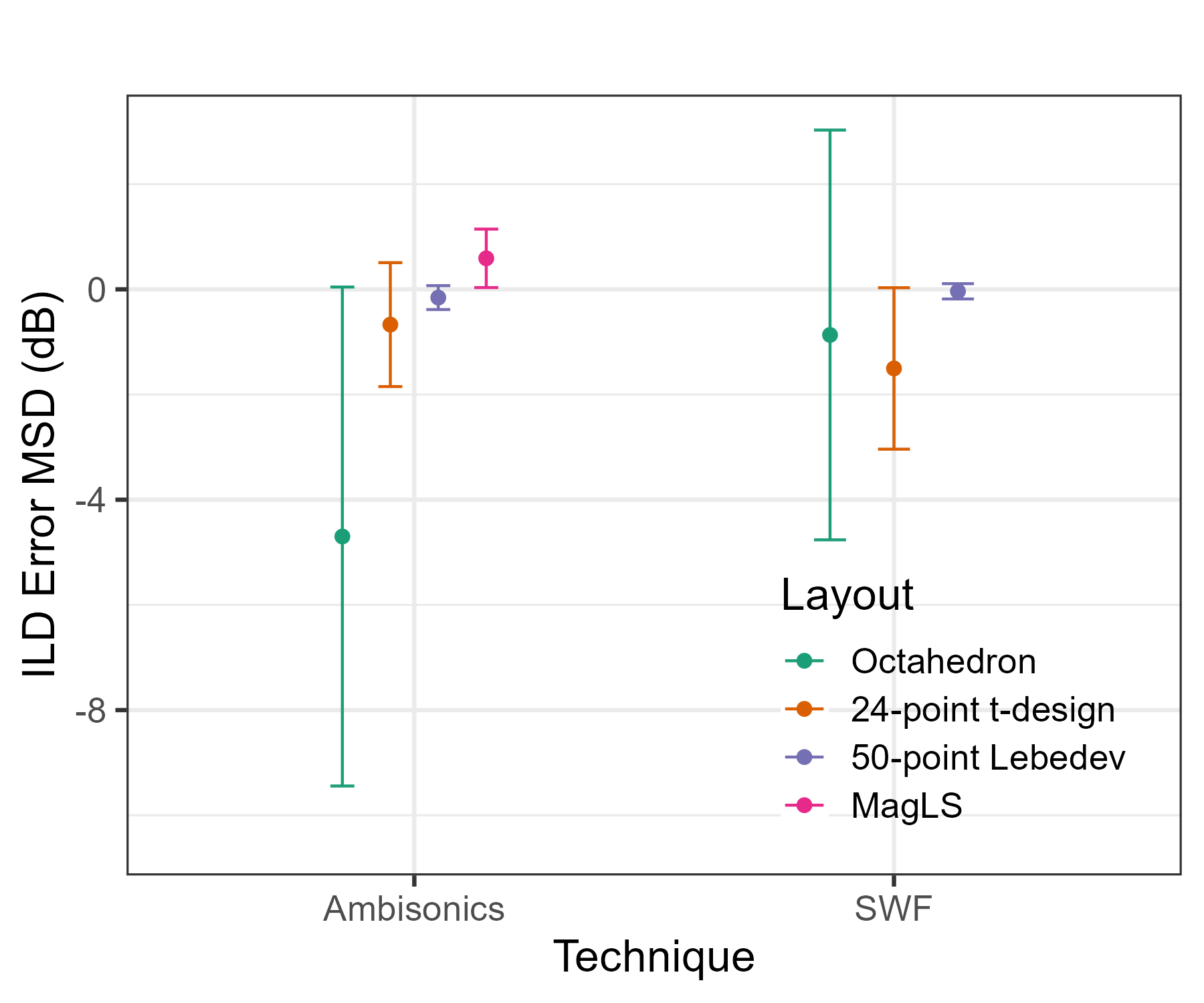}
    \label{fig1b}}
    \quad
    \subfloat[]{\includegraphics[width=2in]{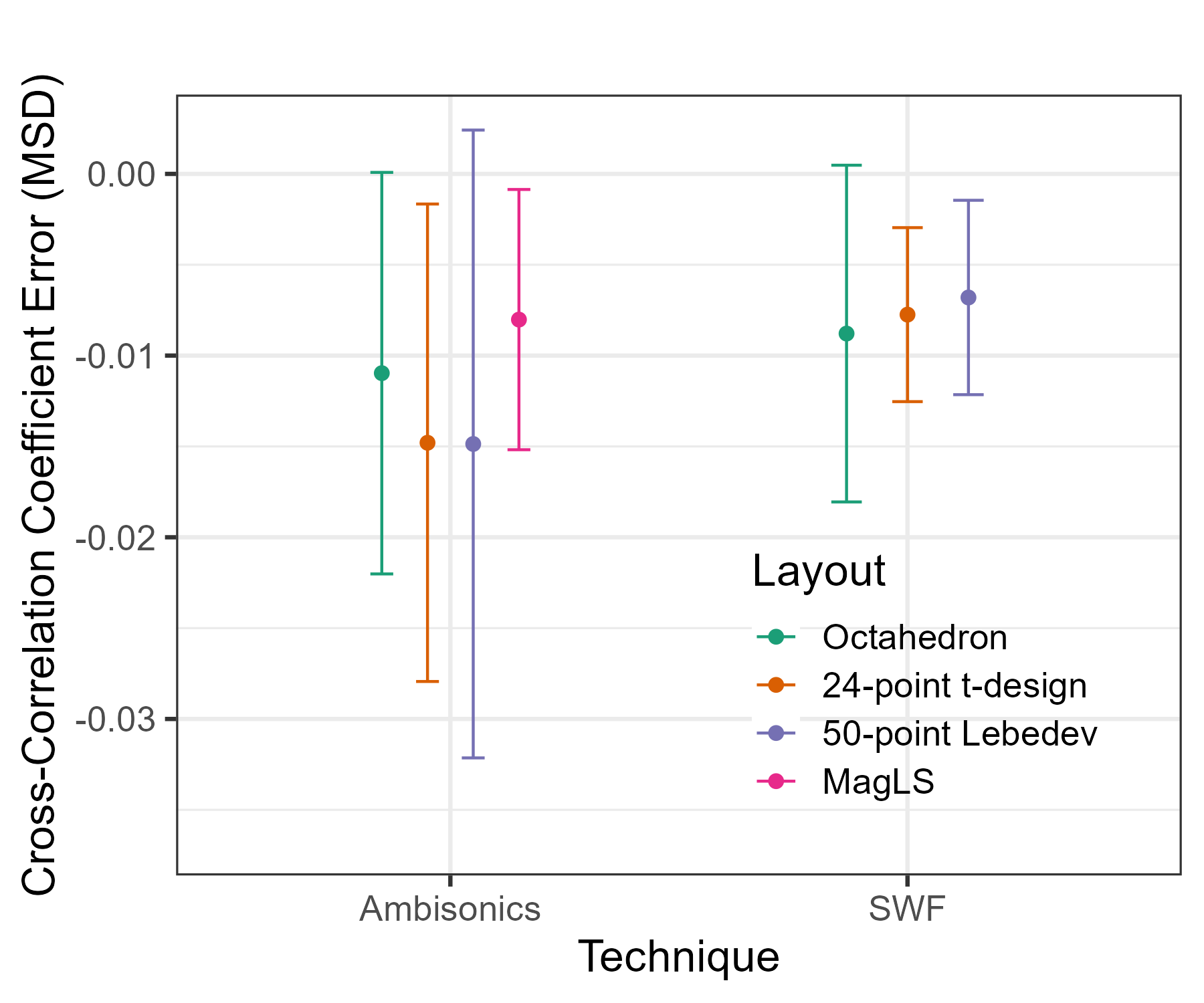}
    \label{fig1c}}
    \subfloat[]{\includegraphics[width=2in]{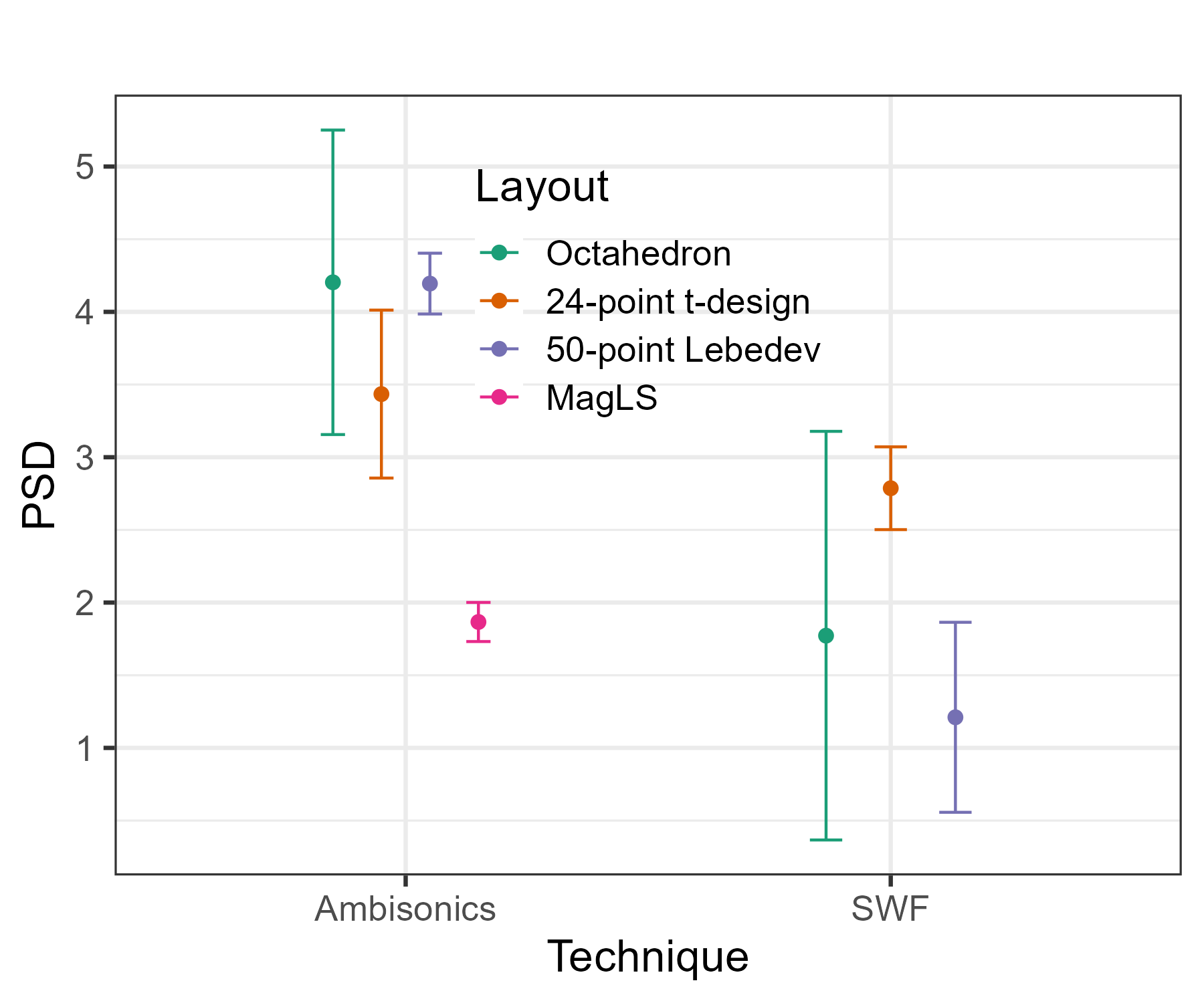}
    \label{fig1d}}
    \caption{(a) ITD, (b) ILD, (c) IACC and (d) PSD errors as MSD with medians and non-parametric 95\% confidence intervals.}
    \label{fig1}
\end{figure*}

Results of the simulations for the ITD error, calculated as the Mean Signed Difference between the system and the reference presented in Fig. \ref{fig1a}. An interesting first observation is that SWF on both 50-point Lebedev and Octahedron has the median ITD error of 0 and this is lower than any other reproduction. This is followed by Ambisonics on 50-point Lebedev which has the median error of 0.00022 milliseconds. It has been reported that JND for ITD with anechoic sound sources is 0.02 microseconds \cite{klockgetherJustNoticeableDifferences2016} and for ITD to effect Minimum Audible Angle (MAA), JND is 10 microseconds \cite{millsMinimumAudibleAngle1958}. Considering these, there are no meaningful difference between Ambisonics on MagLS and SWF on Octahedron and 50-point Lebedev with respect to the reference. The only noticeable ITD error is exhibited by the Ambisonics on Octahedron which has the median value of 125 microseconds. 

Results of ILD error calculations are similarly presented in Fig. \ref{fig1b}. SWF on 50-point Lebedev has the median error of -0.037 dB while Ambisonics on 50-point Lebedev have the median error of -0.157 dB. This is followed by Ambisonics on MagLS with the median error of 0.590 dB. Klockgether and van de Par \cite{klockgetherJustNoticeableDifferences2016} report the JND for ILD with anechoic sound sources as 0.6 dB, while Gao et al. \cite{gaoJNDbasedSpatialParameter2016} shows that JND is dependent on the magnitude of ILD: if the ILD is around the 0 dB, JND is approximately 0.5 dB whereas if the ILD is around the 15 dB, JND increases up to 2 dB. Based on these reports, it can be said that Ambisonics on 50-point Lebedev and SWF on 50-point Lebedev both have ILD errors that are unnoticeable. While Ambisonics on MagLS might have one that is being noticable.  A result worth noting is that SWF on Octahedron has the median error of -0.867 dB with a range between -13.685 dB to 0.028 dB, while Ambisonics on Octahedron has -4.699 Phons with a range between -18.176 dB to  0.456 dB. This tentatively indicates that SWF on Octahedron have comparable ILD performance to reproductions with much higher channel counts such as Ambisonics and SWF on 24-point t-design layout albeit with less stability. Since only other varying factor in these calculations is the panning position, it is likely that this relatively large range of ITD and ILD errors may be due to the performance of reproduction system in relation to the panning position.

Results of the simulations for the IACC error, again as MSD, are presented in Fig. 3.
Overall, median IACC error for SWF on 50-point Lebedev is lower than all of the Ambisonics renderings with value of -0.007. The lowest error for Ambisonics is of MagLS with -0.008. Just Noticeable Difference (JND) for IACC changes with the initial IACC value where an high initial IACC value have higher JND \cite{klockgetherJustNoticeableDifferences2016}. In anechoic environments (where initial IACC generally close to 1), JND is reported to be in the range of 0.03 to 0.07 \cite{durlachInterauralCorrelationDiscrimination1986, gabrielInterauralCorrelationDiscrimination1981}. Especially considering that these models were run for anechoic signals, calculated IACC errors are not meaningful for interpretation.

PSD errors that are presented in Fig. 4, are possible indications of tonal fidelity and elevation localisation, as the localisation on the median plane heavily relies on the spectral content \cite{leeInvestigationPhantomImage2015}. The lowest median PSD is observed with SWF on 50-point Lebedev with 1.211. SWF on Octahedron and Ambisonics on MagLS are following with similar median values of 1.772 and 1.866 while SWF on Octahedron having a wider range. To authors’ knowledge there is no research on the indications of specific PSD values. However, they can be used to compare systems between each other with respect to a reference.
\subsection{Vectorial Analysis}
\subsubsection{Methodology}
Based on Gerzon’s metatheory of auditory localisation \cite{gerzonGeneralMetatheoryAuditory1992} vector analysis entails calculating energy ($rE$) and ($rV$) vectors from the loudspeaker signals and observing their magnitude and directions. It is designed to give information about the reproduction quality in the event of loudspeaker reproduction. A sound reproduction is predicted to be accurate if the magnitude of the energy vector is close to or equal to unity, and the directions of the $rE$ and $rV$ vectors align with the intended virtual source direction. Square sum of all the loudspeaker signals ($E$), is used as a measure of loudness and, for reproduction to be accurate it needs to be as close as possible to the loudness of the actual source \cite{armstrongAmbisonicsUnderstood2021, hellerAmbisonicDecodersNonlinear2010}.  In addition to $rE$, $rV$ vectors and $E$, different spread metrics is proposed by \cite{danielRepresentationChampsAcoustiques2000, epainAmbisonicDecodingConstant2014, frankSourceWidthFrontal2013}. This metrics indicate the perceived source width \cite{zotterAuditoryEventsMultiloudspeaker2019}. Spread metric can also be used as an indication for localisation blur when it is considered as the amount of localisation variance for the source that are localised at the same position \cite{engelAssessingHRTFPreprocessing2022}. Especially in the present study, since the sound source is an anechoic, single source, it is expected to have the narrowest possible width. Therefore, any large amount of width can indicate localisation blur.  All three metrics are calculated using Politis’ library \cite{politisPolarchHigherOrderAmbisonics2025}. Equations used for calculation are as follows:
\begin{align*}
 P=\ \sum_{i\ \in N} s_i && r_V=\frac{1}{P}Re\left(\ \sum_{i\ \in N} s_i\widehat{u_i}\right)
\end{align*}
\begin{align*}
E=\ \sum_{i\ \in N}{(s_is_i^\ast)} && r_E=\frac{1}{E}E\sum_{i\ \in N}{(s_is_i^\ast\widehat{)u_i}}
\end{align*}
Here $P$ is the total acoustic pressure, $E$ is the metric representing total acoustic energy, $s_i$ is the signal of the $i$th loudspeaker and $\widehat{u_i}$ is the unit direction vector in the direction of $i$th loudspeaker of the set of $N$ loudspeakers.
\subsubsection{Results}
\begin{figure*}[!ht]
    \centering
    \subfloat[]{\includegraphics[width=2.15in]{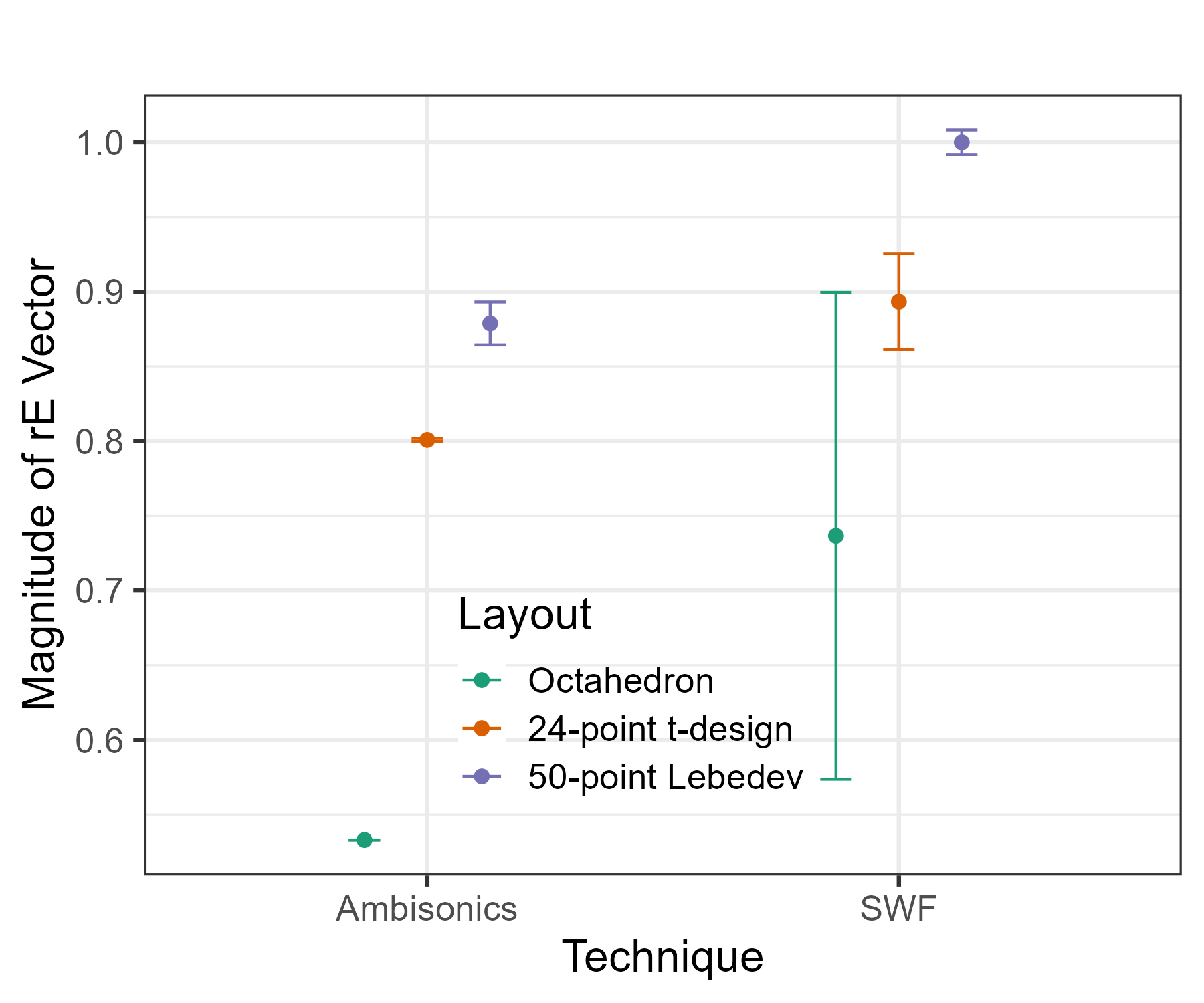}
    \label{fig2a}}
    \subfloat[]{\includegraphics[width=2.15in]{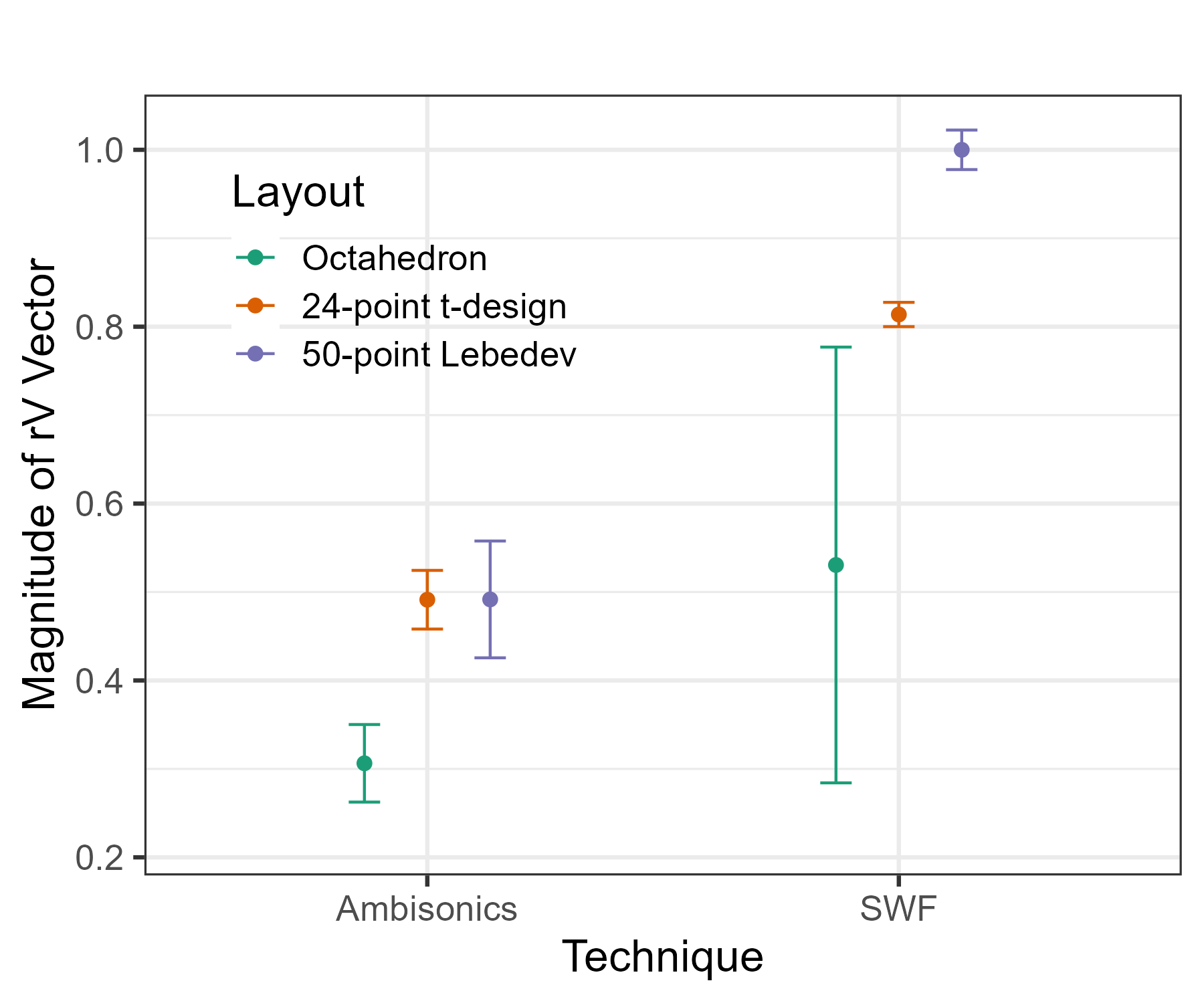}
    \label{fig2b}}
    \subfloat[]{\includegraphics[width=2.15in]{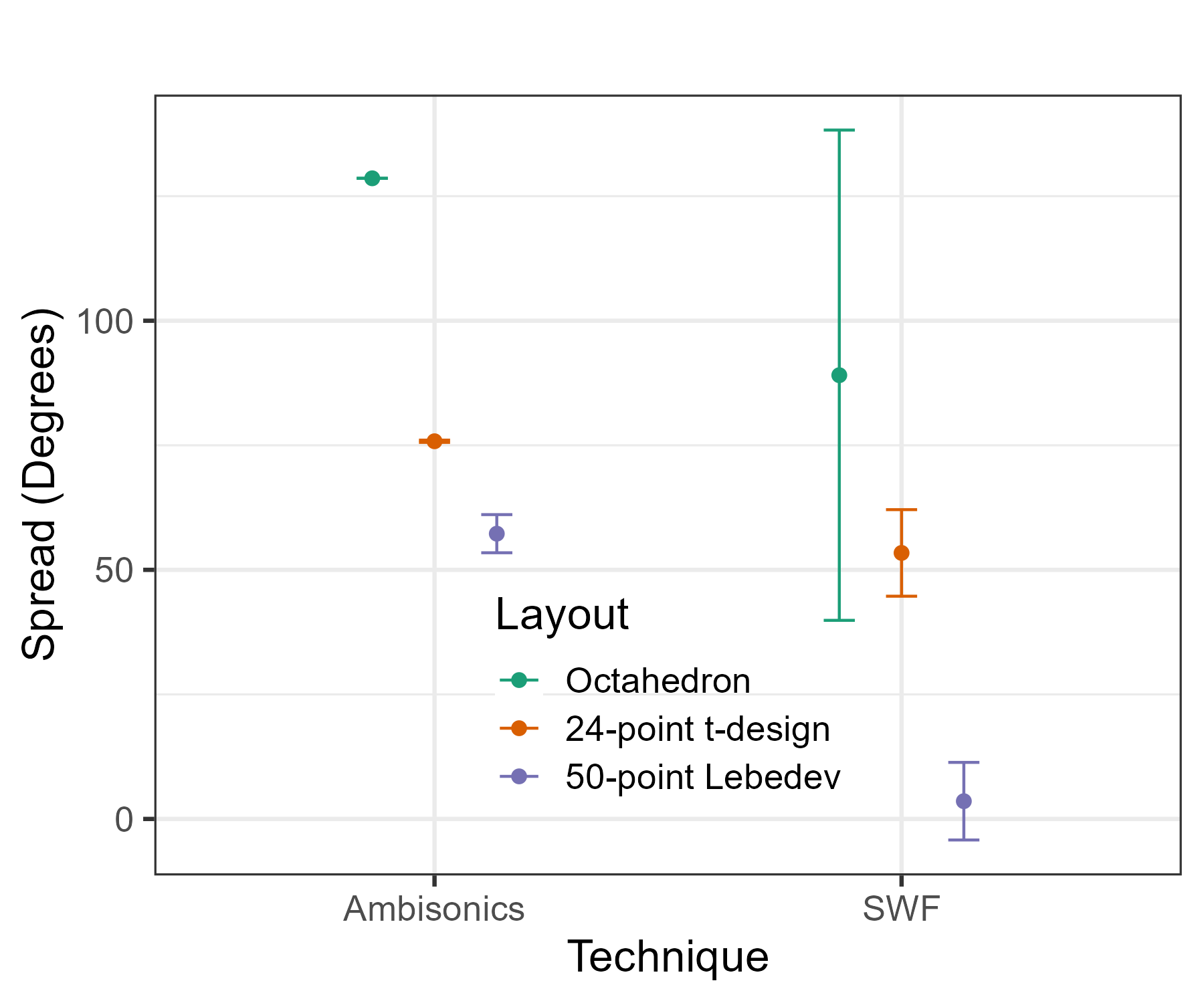}
    \label{fig2c}}
    \caption{Results of vectorial analysis represented as medians and non-parametric 95\% confidence intervals: (a) magnitude of rE vector, (b) magnitude of rV vector (c) spread.}
    \label{fig2}
\end{figure*}

Results of the $rE$ magnitude, $rV$ magnitude and spread calculations are presented in Fig. \ref{fig2a}, \ref{fig2b} and \ref{fig2c}, respectively.
An overall inspection of the plots shows that, for Ambisonics, rE magnitude and rV magnitude increases, while the spread decreases with the Ambisonics order. Moreover, differences between the median values of all three metrics are higher between First-Order Ambisonics (FOA) and Third-Order Ambisonics (3OA) than they are between 3OA and Fifth-Order Ambisonics (5OA). These observations align with previous research on the topic that shows Ambisonics reproduction quality exponentially increases between FOA and 3OA while the rate of increase much lower after the 3OA \cite{leeSpatialTimbralFidelities2019a, zotterXYMSFirstOrder2019}. This is the reason why 3OA has often been taken as the benchmark and it also validates the accuracy of the vectorial simulations in the present study.

Looking at individual plots, median rE magnitude is the highest and unity for SWF on 50-point Lebedev layout and it is followed by SWF on 24-point t-design layout with the median value of 0.893. Theoretically with a single, anechoic source, unity rE magnitude is achievable and SWF achieves it whereas the highest median magnitude for Ambisonics is for 50-point Lebedev layout with the value of 0.879. For the rV magnitudes, SWF on 50-point Lebedev again has the highest median score with the value of 1. However, this time SWF on Octahedron has the median score of 0.531 which is higher than any of the Ambisonics’ albeit having a wider range. Since there is no effect of programme material in vectorial simulations, this range can be attributed to difference of rV magnitudes with different positions which may loosely indicate that SWF on Octahedron has more position dependency compared to others.  Another difference from the rE magnitudes is that median rV magnitudes of 3OA and 5OA are very similar with 0.491 and 0.492 respectively. Looking at the spread metric, SWF on 50-point Lebedev exhibits by far the lowest spread with the median value of 3.567. This is followed by SWF on 24-point t-design which is 53.396 this value is close to the Ambisonics on 50-point Lebedev that has 57.257 and lower than all the other Ambisonics reproductions as well as SWF on Octahedron. Since the used programme material was an anechoic, single source, a very narrow reproduction is expected. Spread metrics indicate that SWF on 50-point Lebedev is the reproduction method that is the closest to that with a very low spread hence the source width.

\section{Listening Test}
\subsection{Methodology}
\label{subsec3}
\subsubsection{Physical Setup}
The listening tests were conducted in The University of Huddersfield, Applied Psychoacoustics Laboratory (APL), critical listening room using binaural reproduction. AKG K702 headphones were fed through Merging HORUS audio interface. Playback level was set to 75 dB SPL (A). This was measured using miniDSP EARs system \cite{MiniDSPEARSUSB} and REW Room Acoustics Software \cite{mulcahyREWRoomEQ}. Loudness consistency between the stimuli was tested by experts from APL. Huddersfield Universal Listening Test Interface Generator (HULTI-GEN) \cite{johnsonHuddersfieldUniversalListening2020} was used as the software interface for the test.
\subsubsection{Stimuli}
For the programme material, 10 seconds long, computer generated pink noise and an approximately 10 seconds long excerpt of anechoic Danish male speech from Music for Archimedes CD \cite{bangMusicArchimedes1992} was used. Input signals were rendered for Ambisonics and SWF as in the same way they were rendered for auditory modelling however, for the listening test, they are headphone equalized using the headphone impulse responses of AKG K702 headphones that were measured by replacing the headphones onto KU100 head 10 times and taking the average of the measurements. After all the processing had performed, stimuli were RMS-I normalised to -25 dB.  
\subsubsection{Participants}
A total of twelve participants took part in the test. They were staff, students, or researchers in audio and music technology at the University of Huddersfield, all with prior experience in spatial audio listening tests. Informed consent was obtained from all of the participants. Of the twelve participants, ten were male and two were female. Their ages ranged from 20 to 48 years, and all reported normal hearing.
\subsubsection{Test Proccedure}
Multiple Stimulus Hidden Reference and Anchor (MUSHRA) procedure \cite{BS1534MethodSubjective} was employed. A modified
scale as in \cite{pawlakSpatialAnalysisSynthesis2024b}, which had verbal labels representing the difference or similarity
to the direct-HRTF rendered reference was used. The scale was a 5-point scale with 0.1 steps that has the scale points and labels: 5 - “The Same”, 4 – “Slightly Different”, 3 – “Different”, 2 – “Very Different”, 1 – “Extremely Different”. 
Before the actual test, a training and familiarisation session conducted for each participant. In these familiarisation
and training sessions participants were informed about the test in detail, tried the test interface, and got familiar with the stimuli.
The test was conducted in two parts: one part was for comparing the stimuli to the reference in terms of tonal fidelity and the other part was for spatial quality. In the tonal quality test, participants were instructed to compare the stimuli and the reference in terms of the spectral content. In the spatial quality test, participants were instructed to compare the stimuli and the reference in terms of location (source incidence) and localisability.  
Each part of the test consisted of 20 trials for the combination of 10 panning positions and 2 different programme material, each trial had 9 tested stimuli: 7 systems under test, 1 hidden anchor and 1 hidden reference. Each part took 40 minutes on average.
\subsection{Results}
\label{subsec4}
\begin{figure}[!ht]
    \centering
    \subfloat[]{\includegraphics[width=2.5in]{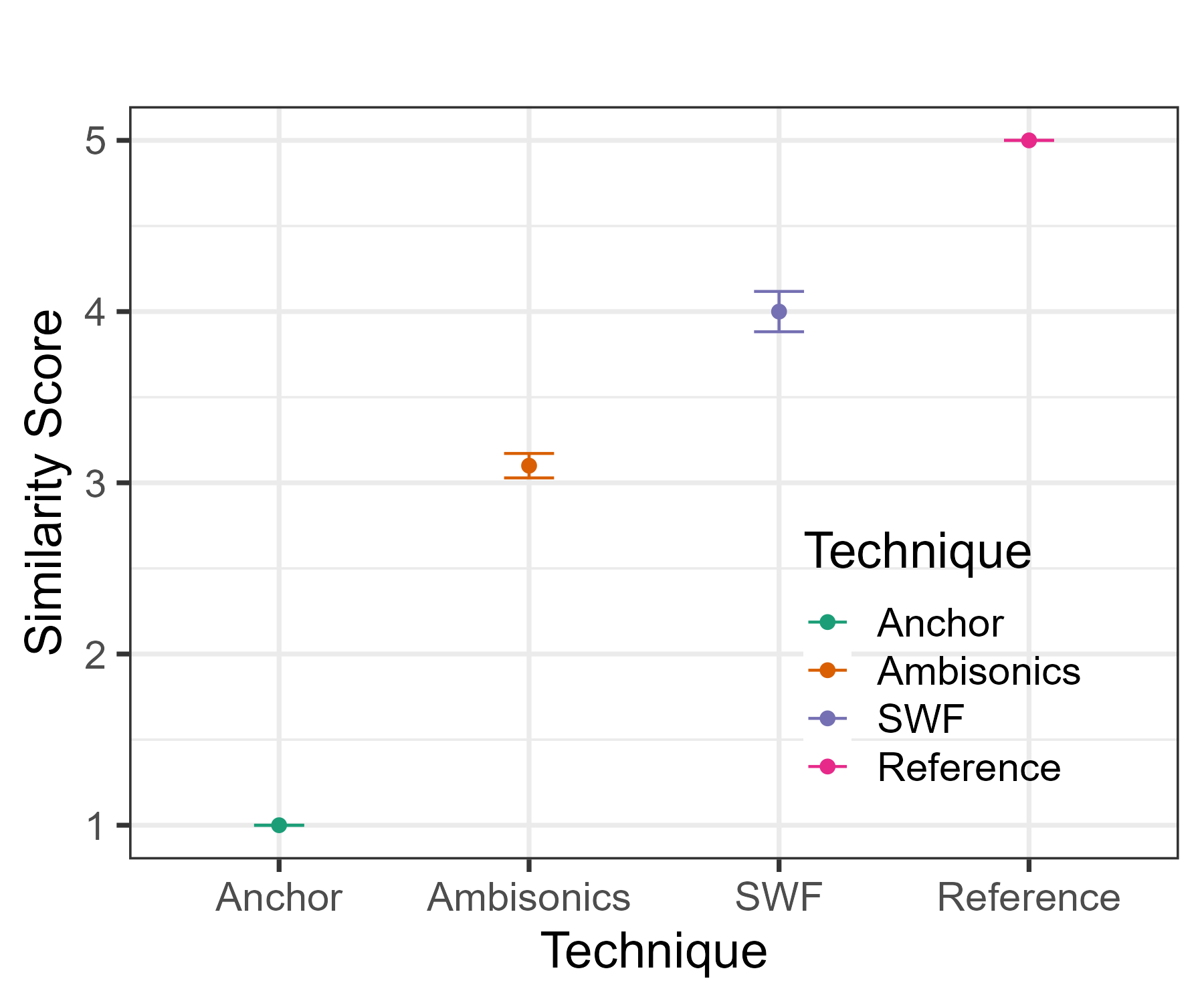}
    \label{fig3a}}\\
    \subfloat[]{\includegraphics[width=2.5in]{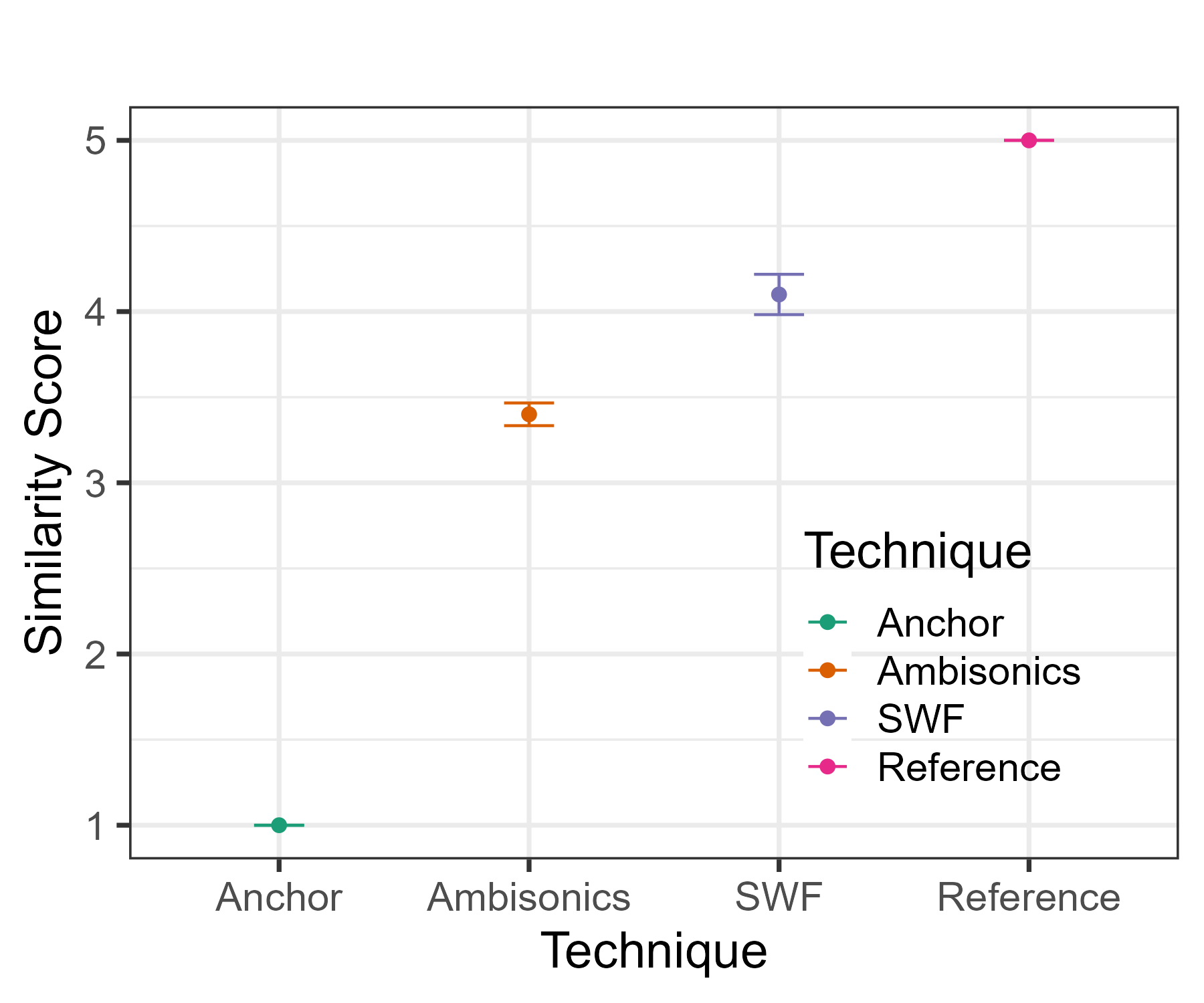}
    \label{fig3b}}
    \caption{Overall comparison of Ambisonics and SWF. Medians and non-parametric 95\% confidence intervals for (a) tonal quality test and (b) spatial quality test.}
    \label{fig3}
\end{figure}

Overall comparison of Ambisonics and SWF with respect to the reference, meaning that all the different layouts, programme
materials, and positions are aggregated, is presented in Fig. \ref{fig3} for both \ref{fig3a} tonal quality test and \ref{fig3b} spatial quality
test. Similarity ratings, given by the participants to each stimulus are represented as notch-edged error bars where the
hinges represent non-parametric 95\% confidence intervals and the points represent the median ratings. Plots indicate that, SWF was rated significantly more similar to the reference than Ambisonics was. This observation is same for both tonal quality and spatial quality. In both aspects, Median rating for SWF is approximately 4, “Slightly Different” while it is 3.4, “Different” for Ambisonics. Since the scale employed for the tests was ordinal, non-parametric methods were used for statistical analysis. Wilcoxon signed-rank tests was employed on the data excluding the ratings for the MagLS layout in order to keep the number of ratings equal for both techniques. Statistical tests showed that the differences between SWF and Ambisonics, both in tonal quality and spatial quality, were significant with large effect size according to Cohen’s criteria ($p = 1.33e-52$, $r = 0.570$) for tonal quality and ($p = 3.44e-54$, $r = 0.580$) for spatial quality. 

\begin{figure*}[!ht]
    \centering
    \subfloat[]{\includegraphics[width=2.5in]{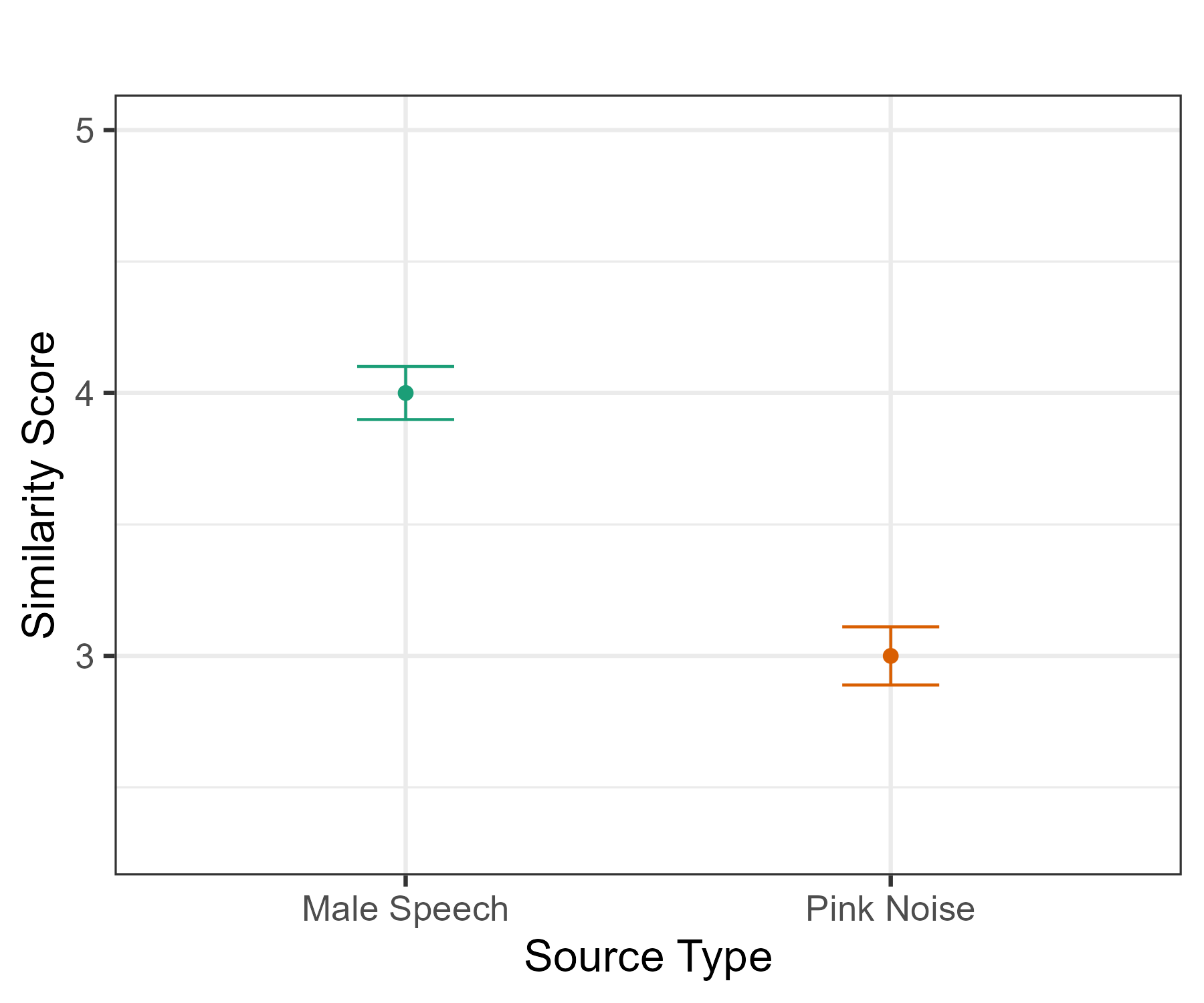}
    \label{fig4a}}
    \subfloat[]{\includegraphics[width=2.5in]{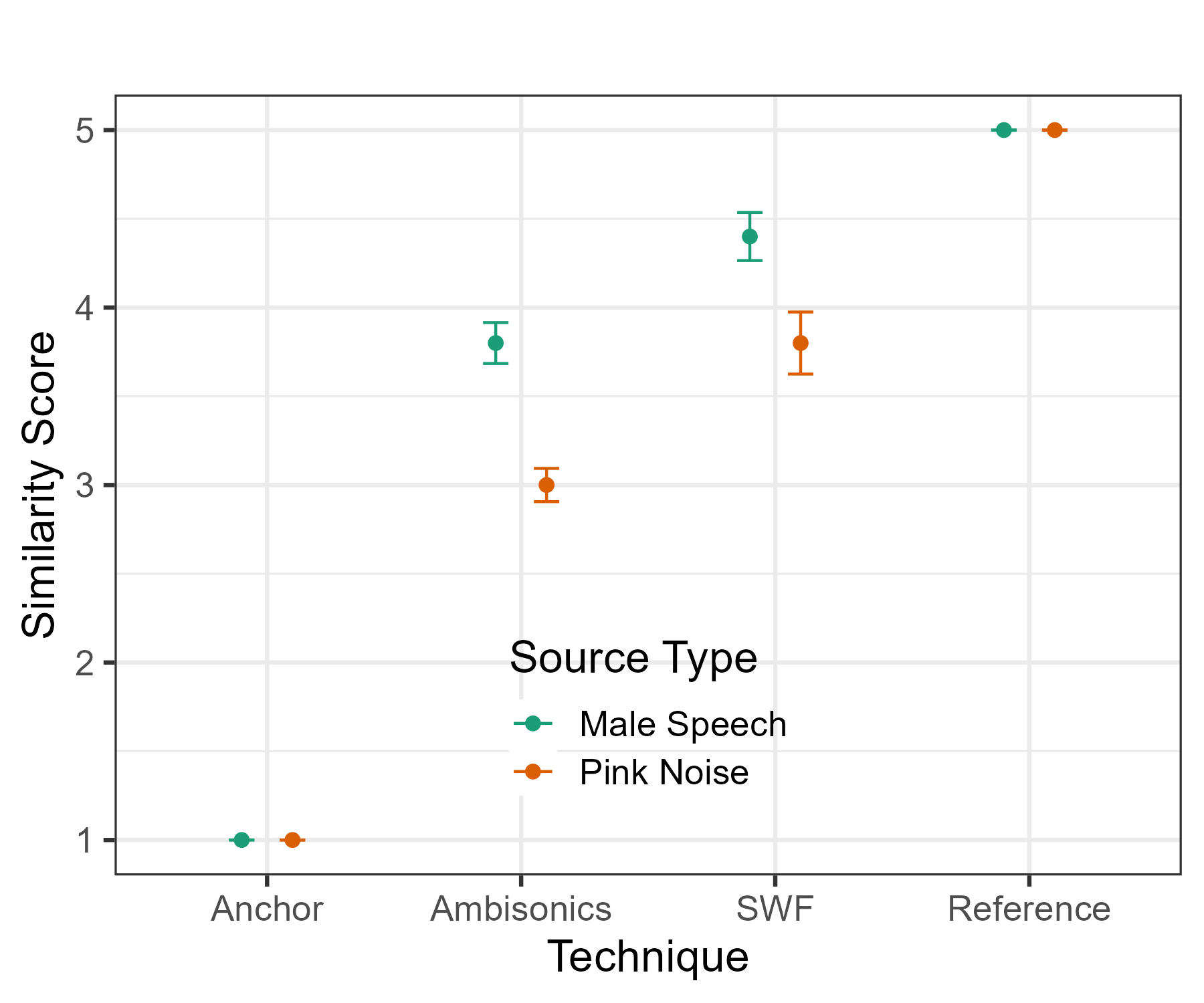}
    \label{fig4b}}
    \quad
    \subfloat[]{\includegraphics[width=2.5in]{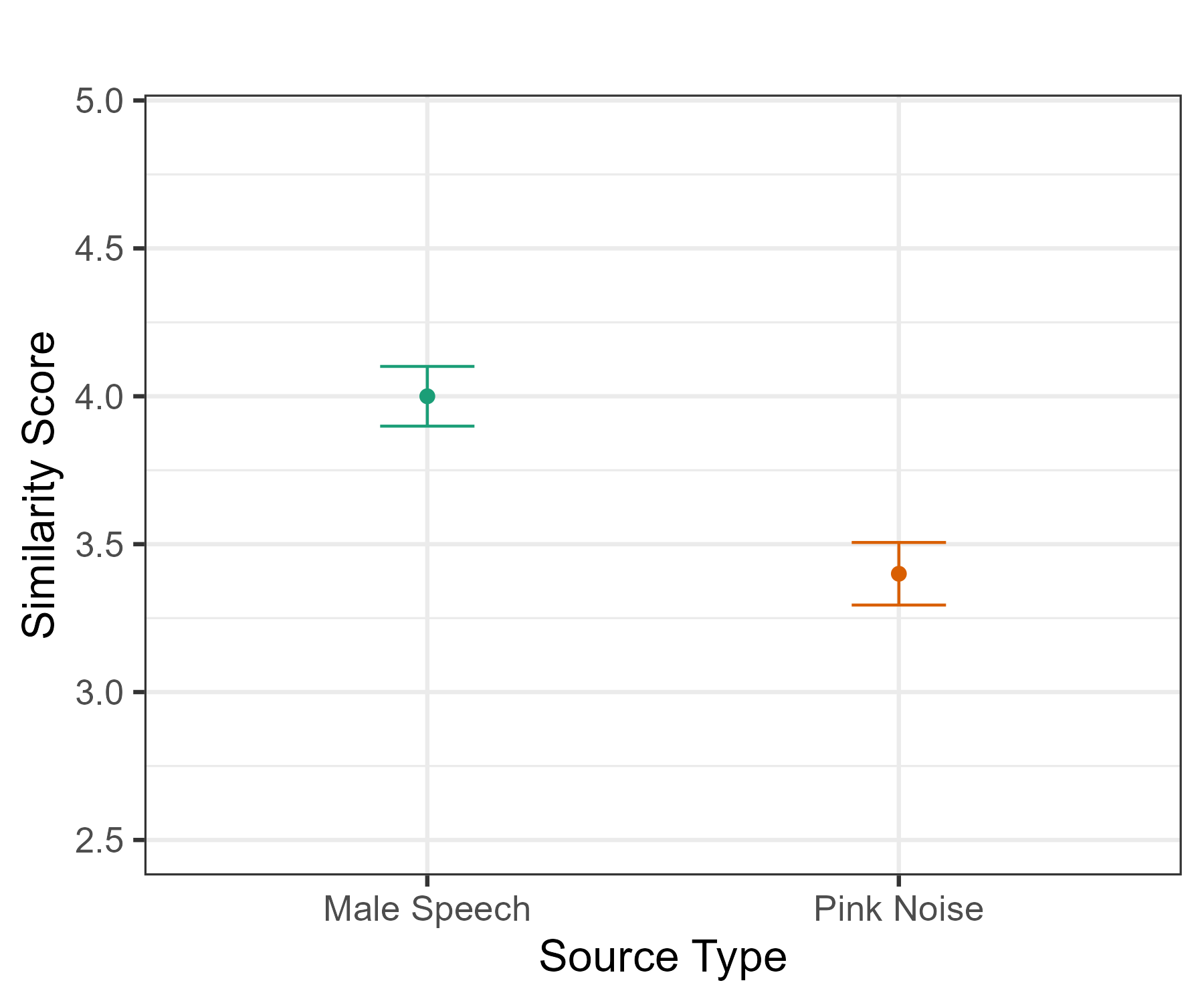}
    \label{fig4c}}
    \subfloat[]{\includegraphics[width=2.5in]{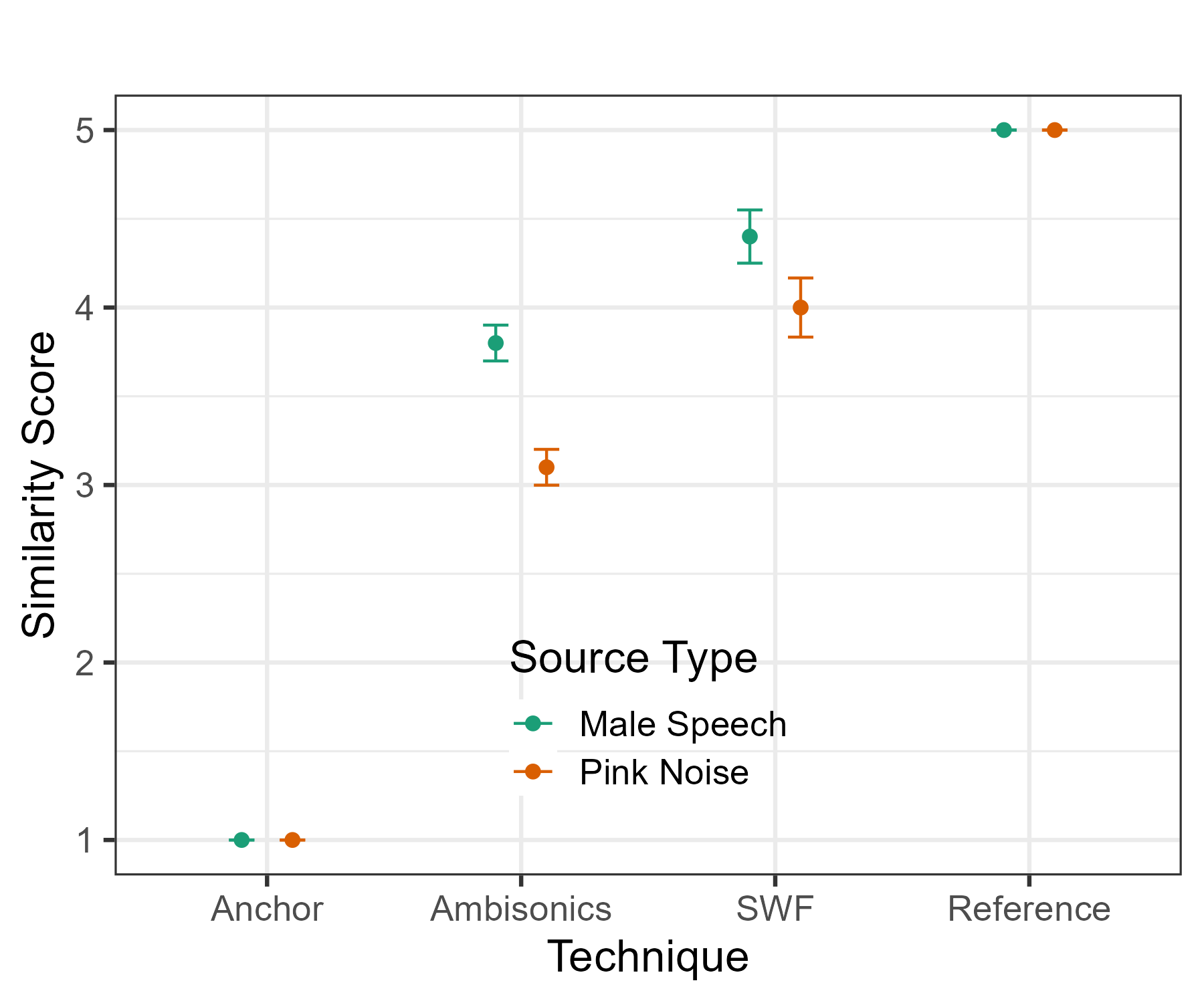}
    \label{fig4d}}
    \caption{Programme material as the main effect. Medians and non-parametric 95\% confidence intervals. (a) All the ratings are aggregated by source for tonal quality, (b) ratings are separated by technique for tonal quality (c) all the ratings are aggregated by source for spatial quality and (d) ratings are separated by technique for spatial quality}
    \label{fig4}
\end{figure*}

Inspection of the programme material as the main effect is presented in the Fig. \ref{fig4} where \ref{fig4a} all techniques, layouts and positions are gathered, and \ref{fig4b} compared within each technique for tonal quality test and \ref{fig4c} all techniques, layouts and positions are gathered, and \ref{fig4d} compared within each technique for spatial quality test.

As indicated by the plots, overall, stimuli with the pink noise as the programme material, rated significantly lower than the stimuli with the male speech excerpt in overall ($p = 5.24e-39$, $r = 0.449$) for tonal quality and ($p = 3.96e-18$, $r = 0.302$) for spatial quality. When the ratings were broken down to different techniques, it can be seen that differences of ratings between the programme materials are higher with Ambisonics with statistical tests showing ($p = 2.00e-25$, $r = 0.548$) compared to SWF ($p= 8.68e-12$, $r = 0.353$) in tonal quality test and ($p = 2.49e-11$, $r = 0.357$) for Ambisonics compared to SWF ($p= 1.40e-6$, $r = 0.246$) in spatial quality test. This tentatively indicates that Ambisonics is more dependent on the programme material when compared to SWF. This also aligns with the results of PSD calculations where SWF has lower overall PSD scores, hence lower spectral difference from reference, compared to the Ambisonics.

\begin{figure}[ht]
    \centering
    \subfloat[]{\includegraphics[width=2.5in]{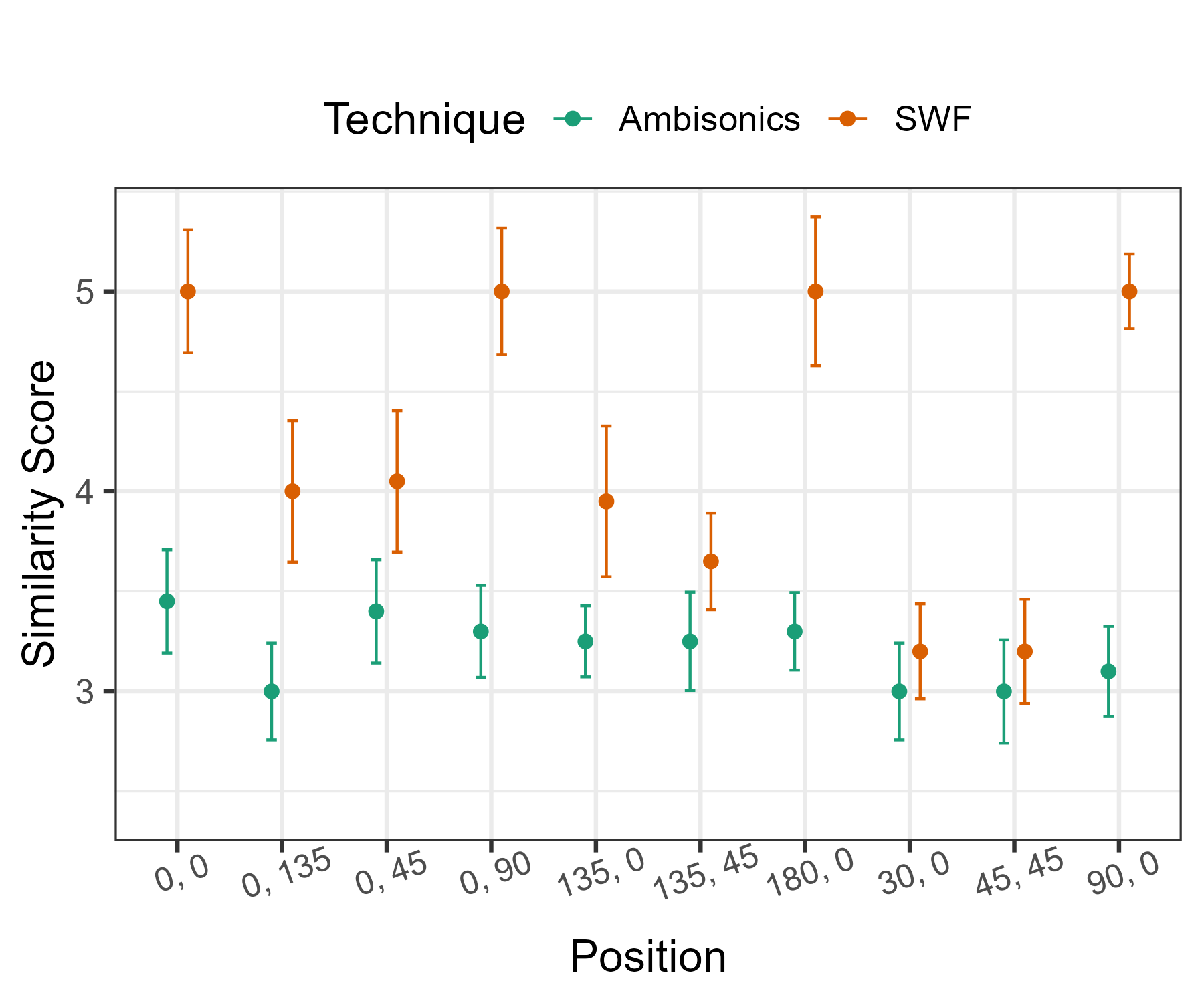}
    \label{fig5a.png}}\\
    \subfloat[]{\includegraphics[width=2.5in]{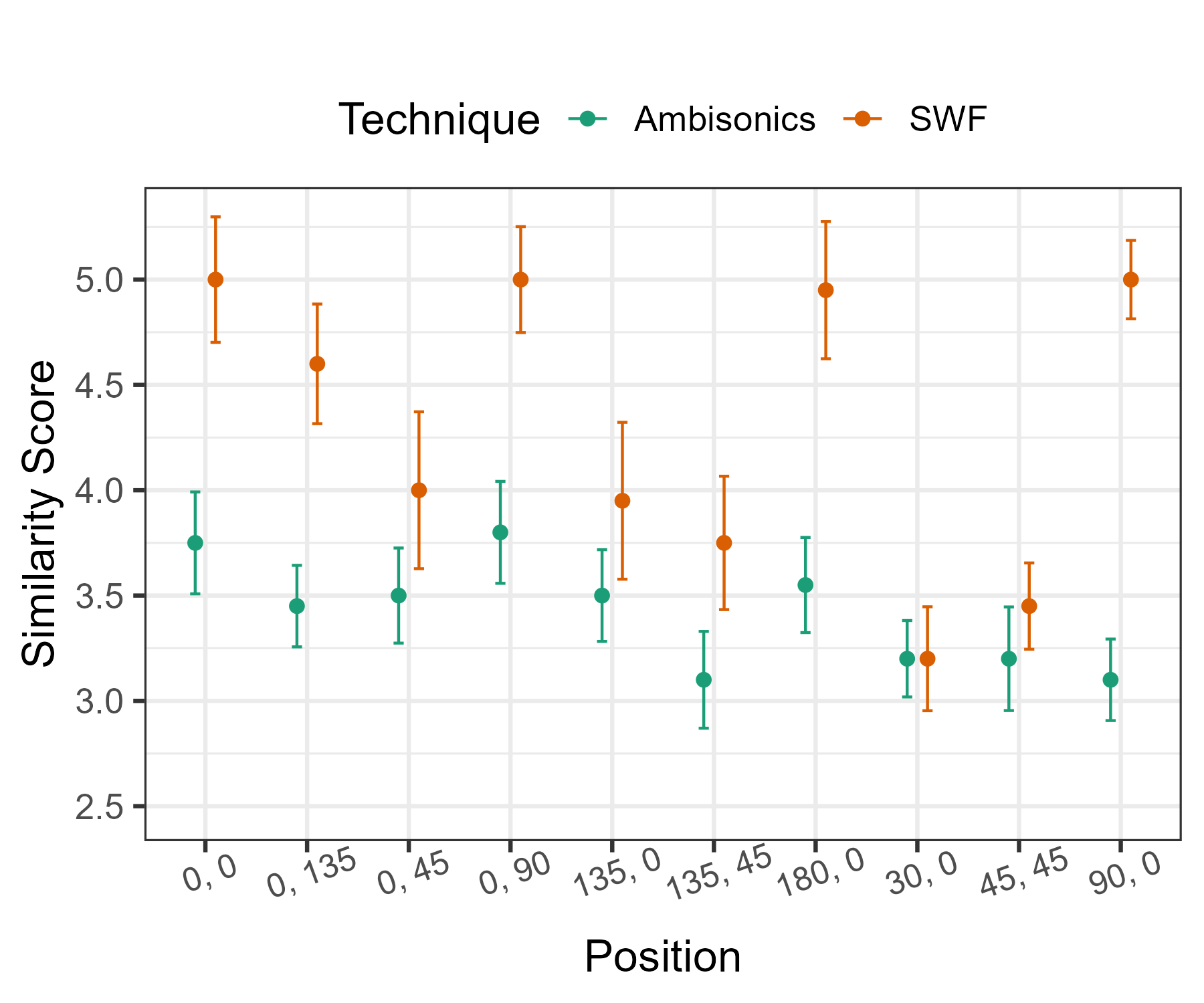}
    \label{fig5b.png}}
    \caption{Position as the main effect (a) tonal quality test and (b) spatial quality test. Medians and non-parametric 95\% confidence intervals.}
    \label{fig5}
\end{figure}

\begin{figure}[t]
    \centering
    \subfloat[]{\includegraphics[width=2.5in]{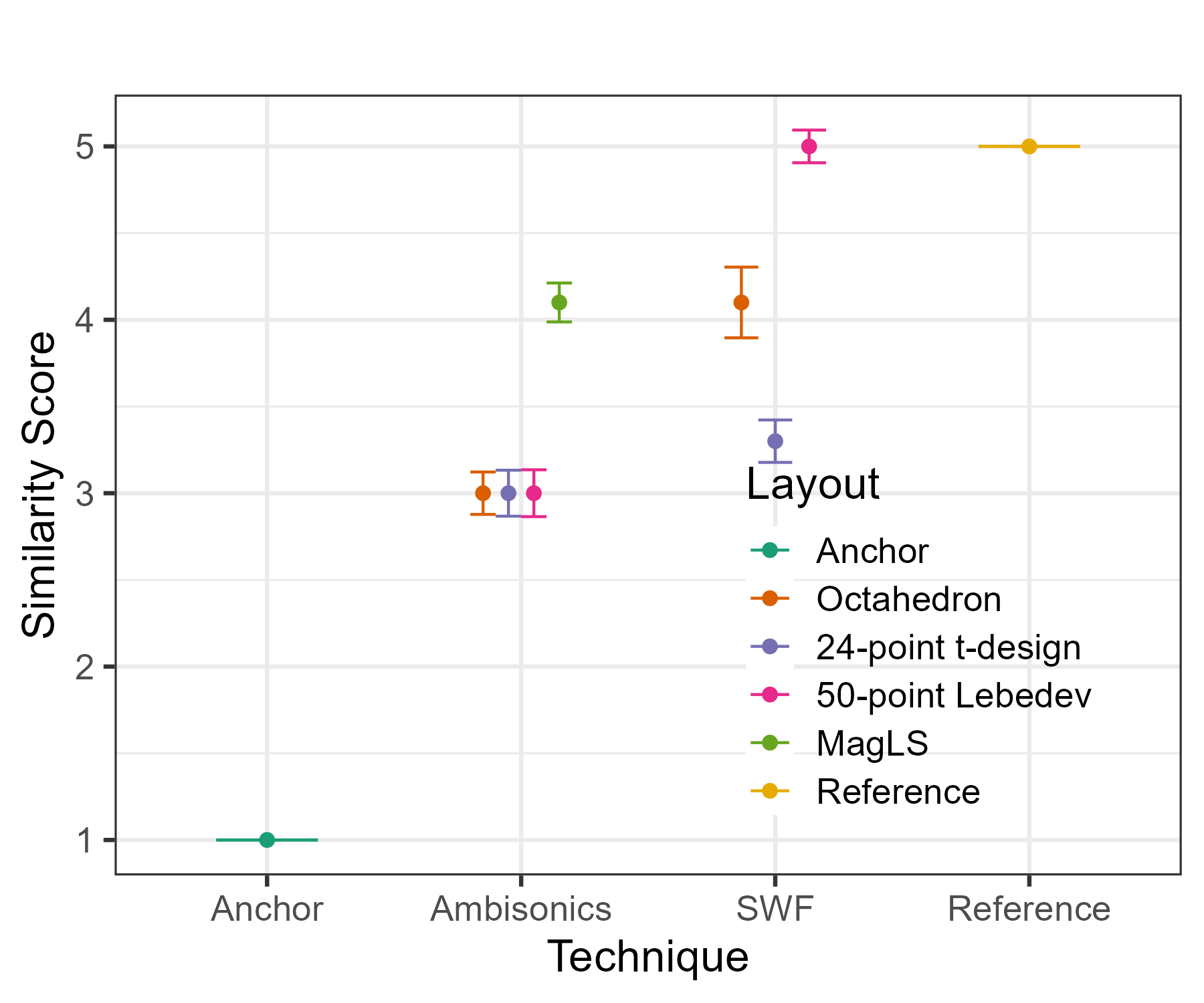}
    \label{fig6a}}\\
    \subfloat[]{\includegraphics[width=2.5in]{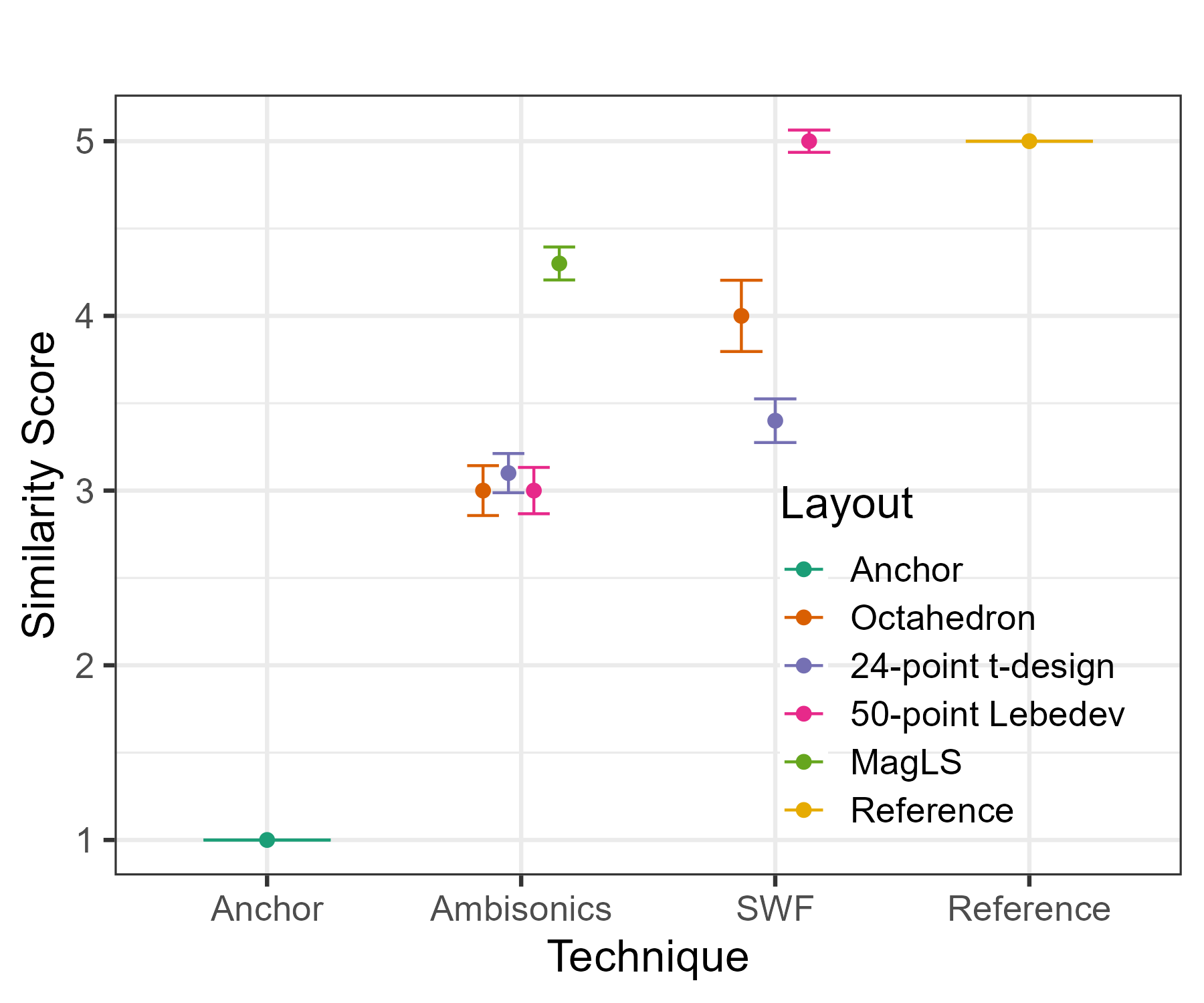}
    \label{fig6b}}
    \caption{Comparison of systems. Medians and non-parametric 95\% confidence intervals for (a) tonal quality test and (b) spatial quality test.}
    \label{fig6}
\end{figure}

Panning position was also investigated as the main effect. Plots are presented in the Fig. \ref{fig5}. It has been seen that some positions especially have lower median ratings than others. Friedman’s test showed that position has significant effect ($\chi^2=82.388$, $df  = 9$, $p = 5.412e-14$) with small effect size ($Kendall's\ W = 0.054$) in tonal quality and again significant effect with  small effect size in spatial quality ($\chi^2=97.679$, $df = 9$, $p = 4.633e-17$, $W = 0.065$). 
When this effect is broken-down to the technique level it has been seen that position has a slightly larger effect in SWF ($W = 0.211$ for tonal quality and $W = 0.164$ for spatial quality) than it has in Ambisonics ($W = 0.028$ for tonal quality and $W = 0.045$ for spatial quality).
For the next level of comparison, techniques on each layout were compared. For the remaining of the text, every unique combination of a technique and a layout is referred to as “system” and it is written as the “Technique/Layout” (e.g. Ambisonics on Octahedron is a system, and it is written as the Ambisonics/Octahedron). Comparison of systems is first done by aggregating all the other factors (i.e. programme material and panning position). Plots showing the ratings with notched-edge, error-bars with the points representing the medians are presented in the Fig. \ref{fig6} for \ref{fig6a} tonal quality test and \ref{fig6b} spatial quality test.
It has been seen that, SWF/50-point Lebedev was the system that is rated most similar to the reference in both tonal quality and spatial quality tests with the median rating of 5, “The Same”. Wilcoxon signed-rank test showed significant difference between the two ($p < 0.001$) in both tests while multiple-comparisons after Friedman’s using the implementation in \cite{MultipleComparisonsFriedman} showed non-significant differences ($p > 0.05$) in spatial quality but significant differences in tonal quality test ($p < 0.05$). SWF/50-point Lebedev is followed by Ambisonics/MagLS which has the median rating of 4.3, corresponding approximately to “Slightly Different”. This is followed by SWF/Octahedron, that has the median rating of 4.1, again, corresponding to “Slightly Different”. Wilcoxon signed-rank test showed that the difference between Ambisonics/MagLS and SWF/Octahedron is significant ($p < 0.005$) for tonal quality and ($p < 0.05$) for spatial quality. This was also the case with multiple comparisons after Friedman’s test, which again showed a significant difference ($p < 0.05$) for both tests. All other Ambisonics systems have the median ratings between 3 and 3.1 for both tests corresponding to “Different” while SWF/24-point t-design have the median rating of around 3.3 and 3.4 which is also around the label “Different”. Results of the pairwise Wilcoxon signed-rank test with Holm p-value correction showed that ratings of SWF/24-point t-design are significantly different from all the other systems other than Ambisonics/24-point t-design. 

\begin{figure*}[!ht]
    \centering
    \includegraphics[width=3.5in]{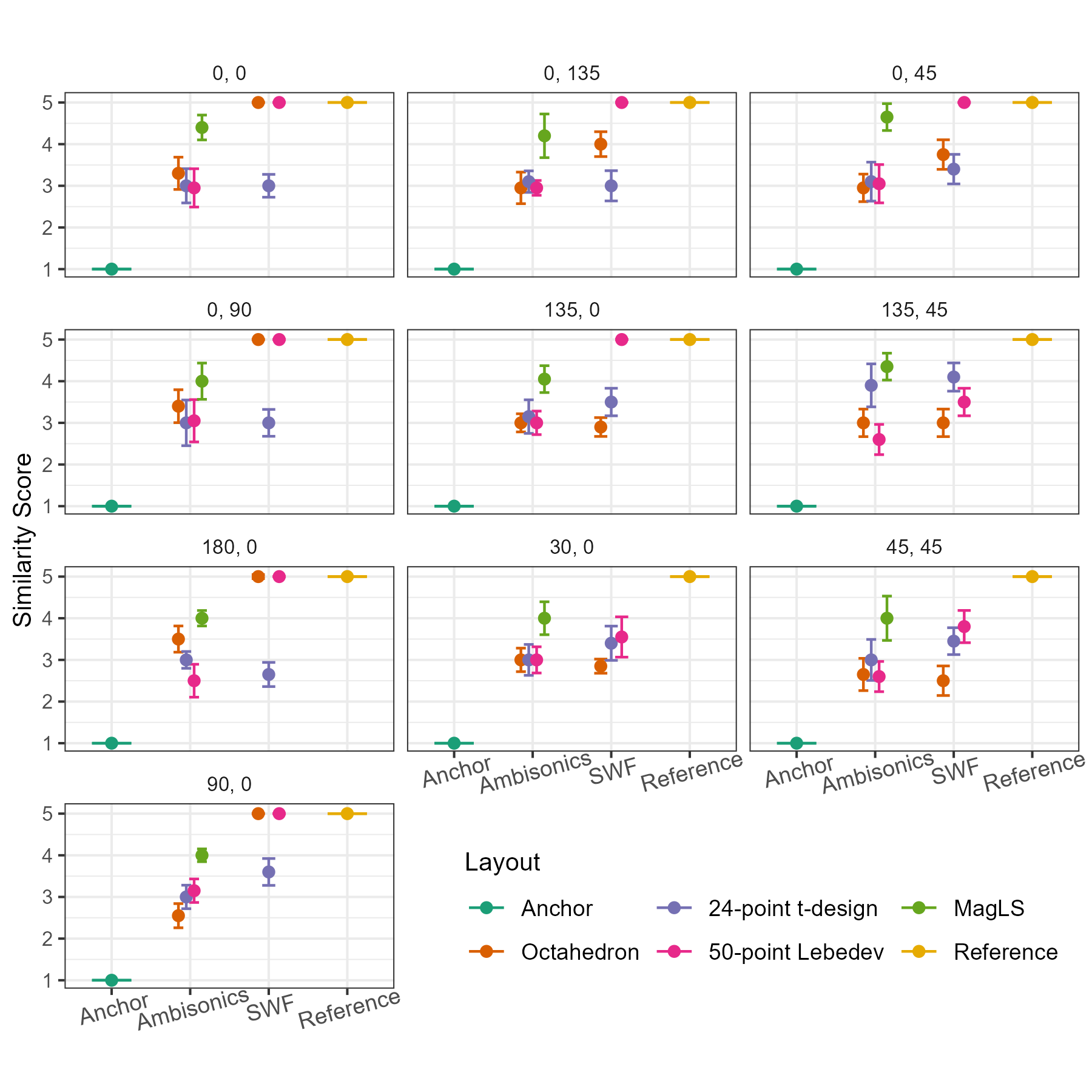}
    \caption{Similarity ratings of each system at each panning position in tonal quality test. Medians and non-parametric 95\% confidence intervals.}
    \label{fig7}
\end{figure*}

\begin{figure*}[!ht]
    \centering
    \includegraphics[width=3.5in]{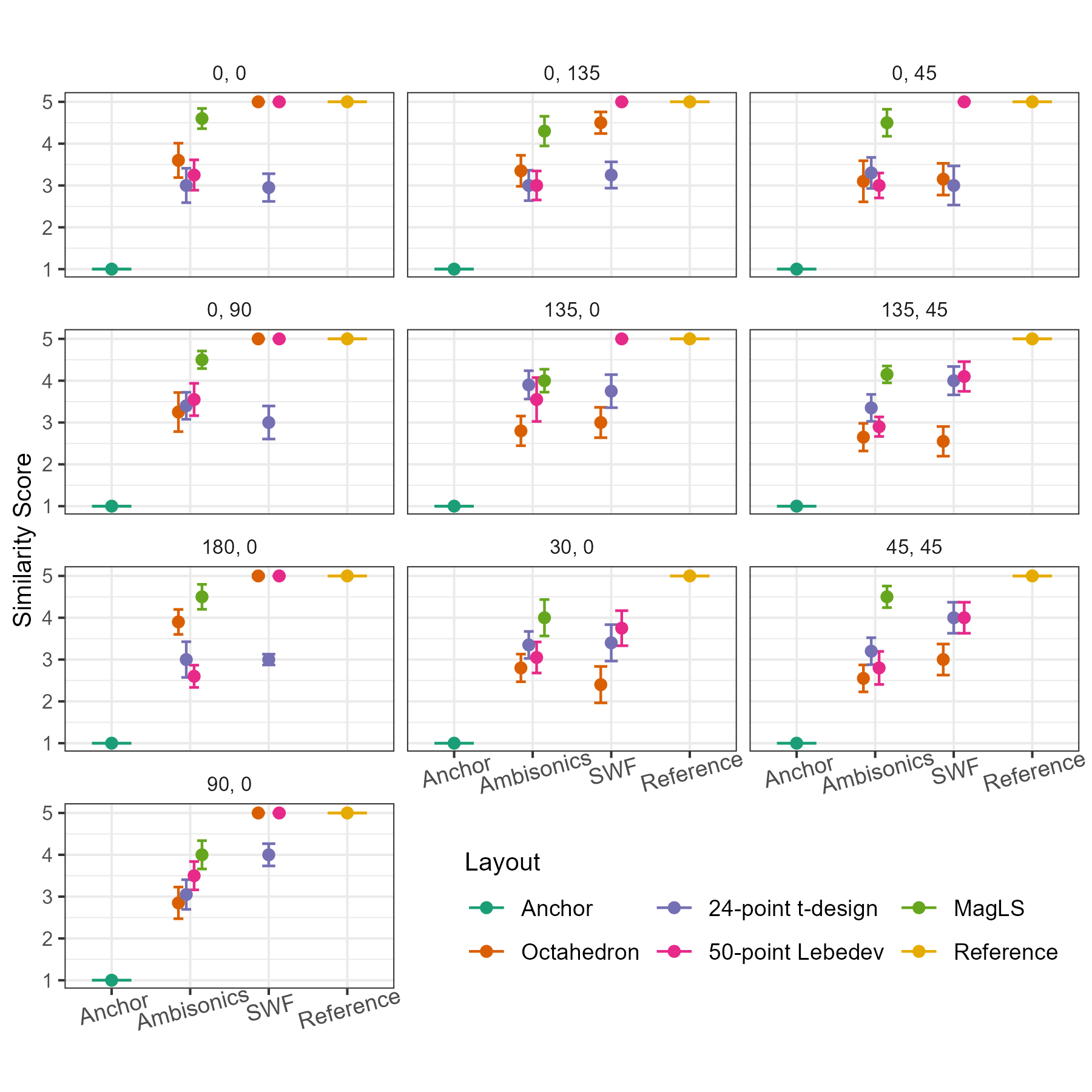}
    \caption{Similarity ratings of each system at each panning position in spatial quality test. Medians and non-parametric 95\% confidence intervals.}
    \label{fig8}
\end{figure*}

Ratings of the systems for each panning position is presented in the \ref{fig7} and \ref{fig8}. Inspecting the plots, it can be seen that some of the systems have varying median ratings depending on the panning position. For example, Ambisonics/MagLS has the median ratings of 4.40 and 4.60 for the tonal and spatial quality tests respectively at the panning position of (0, 0), and likewise 4.65 and 4.50 at (0, 45). On the other hand, SWF/50-point Lebedev has the median rating of 5 for both tests at the panning position of (0,0) while having 3.55 and 3.75 at (30, 0). When Friedman’s test for the effect of position was performed at system level, it has been seen that, for example, position has significant effect on Ambisonics/50-point Lebedev ($p < 0.005$) in both test with the small effect size ($Kendall's\ W = 0.160$ for tonal quality and $Kendall's\ W = 0.180$ for spatial quality), as well as on SWF/50-point Lebedev ($p < 0.001$) in both tests but this time with large effect size ($Kendall's\ W = 0.761$ for tonal quality and $Kendall's\ W = 0.694$ for spatial quality). Overall, position has significant effect on all the systems, other than Anchor and Reference.

\begin{figure*}[!ht]
    \centering
    \subfloat[]{\includegraphics[width=2.75in]{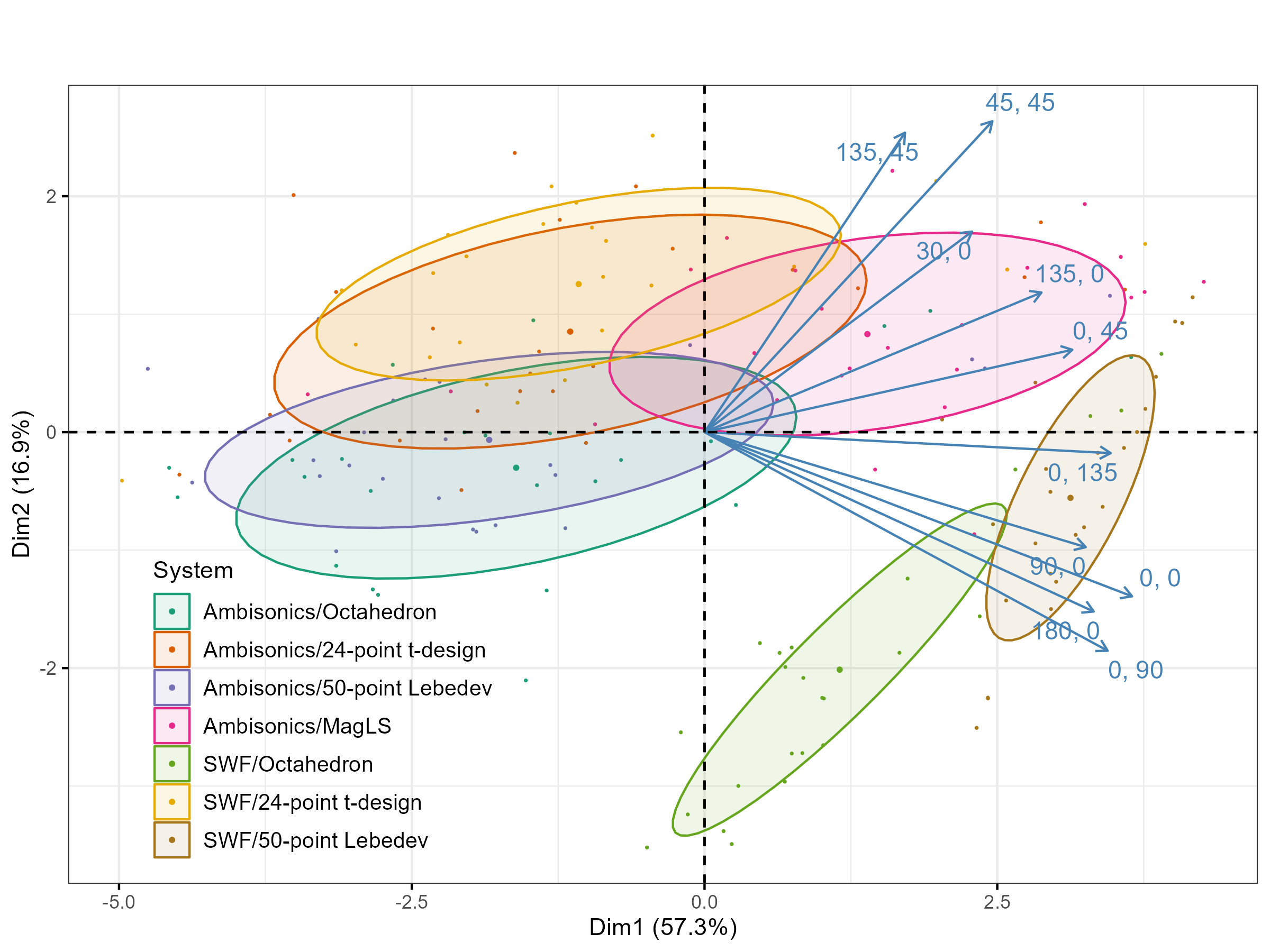}
    \label{fig9a}}
    \subfloat[]{\includegraphics[width=2.75in]{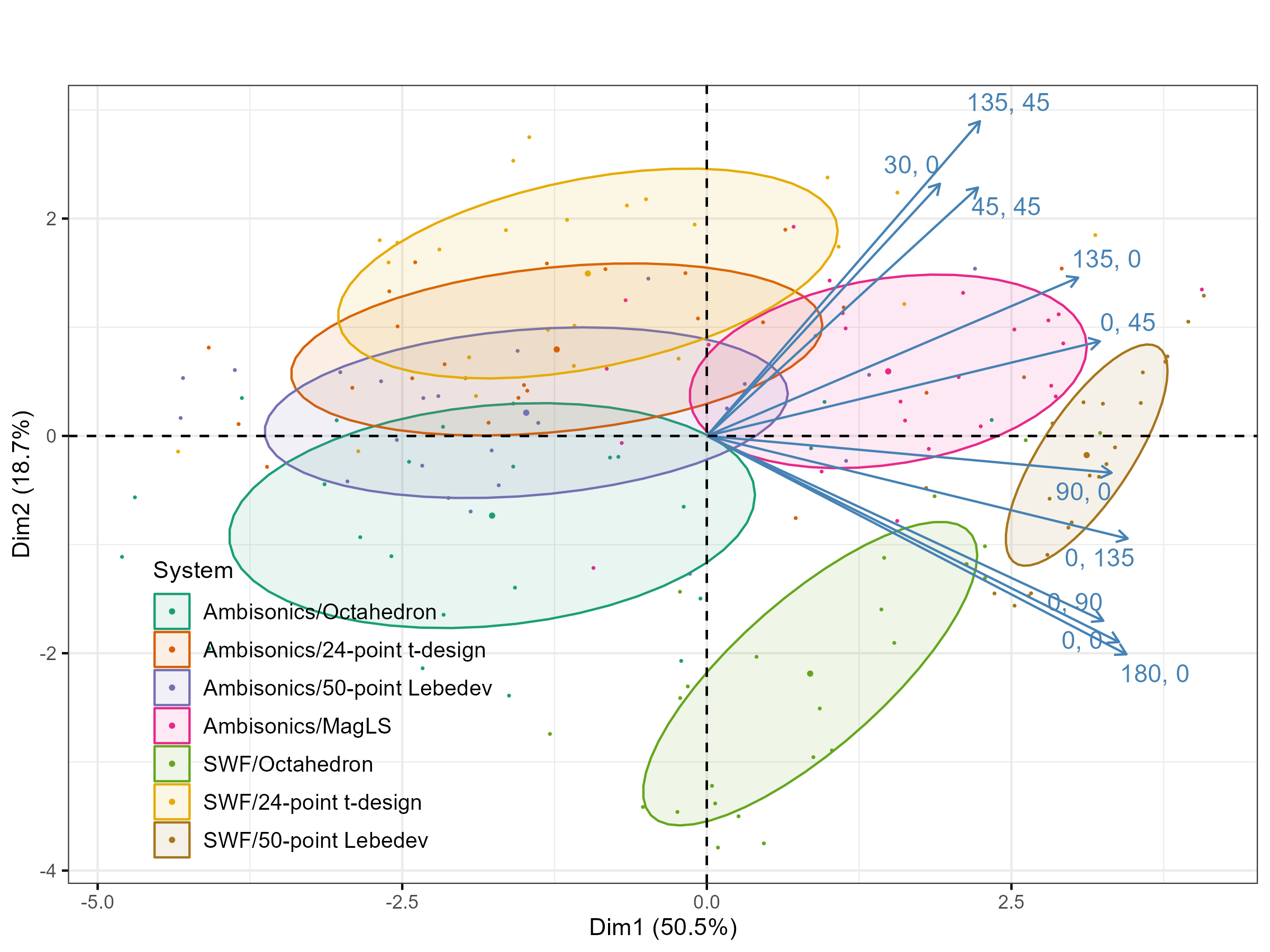}
    \label{fig9b}}
    \caption{PCA of ratings for each system with respect to panning positions (a) tonal quality test and (b) spatial quality test. Arrows representing panning positions as the variable. Scatters representing each participants’ rating for each system. Ellipses representing the systems with bold points as the median values for that system.}\label{fig9}
\end{figure*}

To be able to clearly see the trends and clustering, Principal Component Analysis (PCA) was conducted for reducing the levels of the position. Resulted biplots are presented in the Fig. \ref{fig9}. Using PCA, variances were explained using two dimensions. For tonal quality test these were: Dimension 1 explaining the 57.29\% with the eigenvalue of 5.99 while Dimension 2 explaining 16.86\% with the eigenvalue of 1.76. As for spatial quality test: Dimension 1 explains the 50.51\% with 5.14 and Dimension 2 explains the 18.71 with 1.90. From the biplots, there is a clear grouping of variance between positions (135, 45), (45, 45), (30, 0) and (0, 90), (0, 0), (180, 0). Systems showing similar variance with these positions are SWF/50-point Lebedev, Ambisonics/MagLS, and SWF/Octahedron. To further investigate, multiple comparisons after Friedman’s test as well as pairwise Wilcoxon-signed rank tests were performed to compare positions for each system. As an example, while the differences between the ratings for (135, 45) and (180, 0) were significant according to the Wilcoxon signed-rank test with SWF/50-point Lebedev system ($p < 0.005$), they were non-significant with Ambisonics/50-point Lebedev ($p > 0.05$).

Another noteworthy observation is the similarity between the results of the timbral fidelity and spatial fidelity tests. A Wilcoxon signed-rank test comparing the two for each system reveals statistically significant differences only for the Ambisonics/50-point Lebedev system ($p = 0.004$, $r = 0.18$) and the Anchor ($p = 0.017$, $r = 0.10$), both with small effect sizes. This aligns with findings from \cite{leeSpatialTimbralFidelities2019a} and \cite{pawlakSpatialAnalysisSynthesis2024b}, suggesting a connection between spatial cues and spectral content. As discussed in the Objective Evaluation section, in addition to ITD and ILD, spectral content is especially effective for localisation in the median plane. Moreover, the spectrum of the source signal is also utilized to resolve front-back confusion \cite{leeInvestigationPhantomImage2015}. This connection between spatial and spectral cues is investigated in detail in \cite{leeInvestigationPhantomImage2015}, where the perception of aboveness is shown to be affected by the spectral content of the source signal.

\section{Discussion and Further Study}
Objective analysis results, albeit slight, show a tendency towards SWF, particularly on the 50-point Lebedev grid. Similar trends are also observed with the 6-channel SWF on the Octahedron, where it delivers results comparable to those of 24-channel Third-Order Ambisonics and even 50-channel Fifth-Order Ambisonics across certain evaluation metrics.

Notably, SWF demonstrates better performance in terms of spread and PSD metrics on both the Octahedron and Lebedev grids. These objective findings align well with listening test results. This agreement is expected, given that the spread metric is directly linked to horizontal source width and consequently to localization blur, while the PSD metric relates to spectral fidelity.

However, both the objective and subjective evaluations reveal noteworthy variability particularly for SWF on the Octahedron. Listening tests suggest that this variability largely depends on whether the intended source position coincides with a loudspeaker position. A likely explanation is that, when rendered with SWF, if the source position does not align with a vertex of the densest mesh, tri-linear interpolation is employed. This interpolation process, whether for energy or amplitude, can distort the source image and spectral content.

This issue is not unique to SWF; similar behavior is observed in Vector Base Amplitude Panning (VBAP)\cite{pulkkiVirtualSoundSource1997}, where tri-linear interpolation is also applied. Further examination of loudspeaker signals reveals that SWF typically activates a minimal number of loudspeakers. In cases where the intended source and loudspeaker positions align, only a single loudspeaker may be activated, making SWF effectively behave like direct-to-loudspeaker reproduction. While this extends the sweet spot, it can also lead to inconsistencies in image and timbral quality.

Interestingly, even in panning positions where SWF was rated lower than usual, such as (30°, 0°) or (45°, 45°), it still performed on par with or better than Ambisonics. For instance, at (30°, 0°) on the 50-point Lebedev grid, SWF activates only 16 loudspeakers, compared to at least 42 in Ambisonics for the same layout. This supports the idea that SWF offers a larger sweet-spot area.

Moreover, as discussed in \cite{zotterAuditoryEventsMultiloudspeaker2019}, a better approach may involve activating a small, consistent number of loudspeakers to ensure more uniform performance for every panning position. Instead of activating a different number of loudspeakers depending on the panning position, keeping the number of loudspeakers constant but small allows source spread and inter-channel crosstalk to remain consistent at each virtual source position. This, in turn, leads to reduced localisation blur and spectral distortion. Although this approach may result in greater source spread and inter-channel crosstalk compared to using a single active loudspeaker, it ensures consistency across all panning positions, thereby causing less distraction for the listener. This principle is also reflected in the design philosophy behind Multiple-Direction Amplitude Panning (MDAP) \cite{pulkkiUniformSpreadingAmplitude1999a}, developed as an enhancement to VBAP.

To improve SWF’s consistency across panning positions, one potential solution is to implement a different interpolation method, such as cubic spline interpolation. This could be combined with more advanced subdivision schemes, as proposed in \cite{chambodutWaveletFramesAlternative2005}. Alternatively, exploring different mother wavelets might be beneficial. The original SWF study used the lifting scheme with lazy wavelets, which are known not to produce smooth wavelets, a limitation that could be addressed with a more refined wavelet design.
In the preliminary studies it has been seen that to be able to render a compact microphone array such as tetrahedral or spherical (e.g. Eigenmike em32), a set of wavelet filters should be designed that matches the capsule positions and also be in the same subdivision sequence as the reproduction layout. This is another limiting factor of not necessarily spherical wavelets but Spherical Wavelet Framework.

As it exists, SWF is more in the line of a scaling solution for an object like panning function. It shows promising performance for applications such as substantial number of sound sources (e.g. more than 128) can be encoded to a satisfactorily dense mesh (e.g. 50 or 60 vertices) and then decoded to different irregular layouts. Another one might be that custom decoders for unconventional layouts can be designed that then can be fed with the channels that are acquired from any other format. Besides these SWF provides evidence that spherical wavelets themselves as localised basis functions exhibits promising behaviour for spatial audio compared to spherical harmonics. 

Based on this evidence present authors will study other SWT approaches by Wiaux \cite{wiauxExactReconstructionDirectional2008} and Freeden and Windhauser \cite{freedenSphericalWaveletTransform1996} as these will provide desired properties for a complete field-based spatial audio format. These main properties are (i) steerability where a function at a continuous position on the sphere can be represented by a linear combination of limited set of basis functions, which alleviate the need for a separate interpolation step and allow for encoding of any capsule configuration and (ii) scale discretization which will allow for treating different frequency components differently to a certain extent as in dual-decoding scheme for Ambisonics.

\section{Conclusion}
This study investigated recently proposed Spherical Wavelet Framework through comparison with Ambisonics. Objective analysis consists of IACC, ITD, ILD, PSD metrics was conducted. Error or distance is obtained with respect to the reference that is the direct HRTF convolution. Further objective evaluation was performed using loudspeaker reproduction vectors.  Although slight, they showed tendency towards the SWF being closer to the reference. This is followed by a listening test using MUSHRA procedure and a rating scale that is modified to reflect similarity to the reference. Results show that SWF is perceptually more similar to the reference than Ambisonics is. However, the results revealed that localisation accuracy for certain positions is significantly lower than for others, due to the lifting scheme used in wavelet design not generating smooth wavelets. Moreover, second-generation wavelets with a geometric multi-resolution algorithm do not belong to a rotation group. Because of this, to the authors' knowledge, there is no easy-to-implement method for rotating wavelet coefficients. These shortcomings prevent SWF from being a complete sound-field-based spatial audio format. Nevertheless, the presented performance of SWF can serve as proof that, by using Spherical Wavelets for spatial audio, equal or better localisation accuracy and spectral fidelity can be achieved while requiring fewer channels than Ambisonics.

\section*{Acknowledgments}
This project was funded by the University of Huddersfield. The authors thank everyone who participated in the listening test.

\bibliographystyle{IEEEtran}
\bibliography{references.bib}

\vfill

\end{document}